\newcommand{\ignore}[1]{}

\documentclass[manuscript, screen]{acmart}

\usepackage{url}
\usepackage{multirow}
\usepackage{subfigure}
\usepackage{epsfig}
\usepackage{graphicx}
\usepackage{amsmath}
\usepackage{color}
\usepackage{setspace}
\usepackage{hhline}   
\usepackage{algorithm}
\usepackage{algpseudocode}
\usepackage{flushend}
\usepackage{url}
\usepackage[normalem]{ulem}
\usepackage{graphics}
\usepackage{tikz}
\usepackage{array}
\usepackage{animate}
\usepackage{titlesec}
\usepackage{balance}
\usepackage{enumitem}
\usepackage{booktabs}
\usepackage{subcaption}
\usepackage{pifont}
\usepackage{comment}
\usepackage{hyperref}

\titlespacing*{\section}{0pt}{2pt}{2pt}
\titlespacing*{\subsection}{0pt}{2pt}{2pt}
\titlespacing*{\subsubsection}{0pt}{2pt}{2pt}

\begin{document}

\captionsetup[figure]{font=small,labelfont=small}

\title{Evaluating Computing Platforms for Sustainability: A Comparative Analysis of FPGAs against ASICs, GPUs, and CPUs}

\author{Chetan Choppali Sudarshan}
\affiliation{%
  \institution{Arizona State University}
  \city{Tempe}
  \state{Arizona}
  \country{USA}
}

\author{Aman Arora}
\affiliation{%
  \institution{Arizona State University}
  \city{Tempe}
  \state{Arizona}
  \country{USA}
}

\author{Vidya A. Chhabria}
\affiliation{%
  \institution{Arizona State University}
  \city{Tempe}
  \state{Arizona}
  \country{USA}
}

\renewcommand{\shortauthors}{Sudarshan et al.}
\renewcommand{\shorttitle}{Evaluating Computing Platforms for Sustainability}


\begin{abstract}
\noindent 
Climate change concerns emphasize the need for sustainable computing. Modeling the carbon footprint (CFP), including operational and embodied CFP from semiconductor use, manufacture and design, is essential. Field programmable gate arrays (FPGAs) stand out as promising platforms due to their reconfigurability across various applications, enabling the amortization of embodied CFP across multiple applications. This paper introduces GreenFPGA, a tool estimating the total CFP of FPGAs over their lifespan, considering uncertainties in CFP modeling. It accounts for CFP during design, manufacturing, reconfigurability (reuse), operation, disposal, testing, and recycling. 
GreenFPGA identifies deployment regimes in which FPGAs can be more sustainable than ASICs, GPUs, and CPUs under the modeled iso-performance assumptions.
Experimental results highlight the importance of analyzing applications across different computing platforms to assess their CFP while varying parameters such as application type, lifetime, usage time,  and volume impact their total CFP. Across the evaluated pairwise iso-performance case studies with ASICs, GPUs, and CPUs, FPGAs can be more sustainable under specific deployment regimes involving frequently changing, diverse workloads and low-volume applications.
\end{abstract}

\maketitle

\section{Introduction} 
\label{sec:intro}
\noindent
From microchips to data centers, computing carries a significant carbon footprint (CFP). Rising computational demand, fueled by applications such as artificial intelligence (AI)~\cite{green-ml-1}, has made the information and communication technology (ICT) sector responsible for 2.1\%–3.9\% of global CFP \cite{Freitag2021TheRC}. While the semiconductor industry has focused on making chips smaller, faster, and more energy-efficient to reduce operational CFP, the ecological impact of design, manufacturing, and disposal—embodied CFP—is equally critical \cite{act}.

Previous studies~\cite{act, sudarshan2023ecochip} highlight the significant impact of embodied carbon in data centers and edge devices, offering tools to calculate CFP for monolithic and chiplet-based systems. However, they overlook field-programmable gate arrays (FPGAs) and reconfigurable platforms. While ECO-CHIP~\cite{sudarshan2023ecochip, ecochip-github} applies the 3R concept (reduce-reuse-recycle) by reusing chiplets, this paper introduces a fourth “R,” focusing on reconfigurable computing to further reduce CFP~\cite{fpga-sustainability}.

When considering the spectrum of processing platforms, including FPGAs, graphics processing units (GPUs), central processing units (CPUs), and application-specific integrated circuits (ASICs), FPGAs possess a high degree of reconfigurability, allowing developers to tailor hardware to the demands of specific applications precisely. CPUs, while inherently more general-purpose than GPUs, retain a degree of flexibility that enables their deployment across a broader range of tasks. However, the differing levels of flexibility seen in GPUs, CPUs, and FPGAs come at the cost of varying energy efficiency, making some platforms less efficient than ASICs. Analyzing the full lifecycle CFP of these devices from a sustainability standpoint is crucial for identifying the most suitable computing platform for an application.


\begin{figure}[t]
\centering
\includegraphics[width=0.65\linewidth]{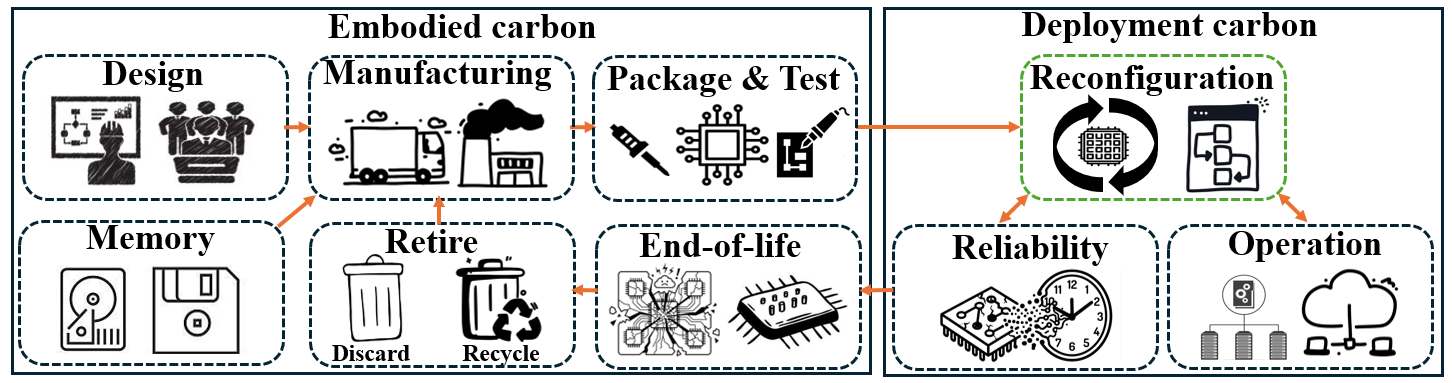}
\caption{Lifecycle of a computing platform: Highlighting the detailed embodied and operational CFP from cradle to grave.}
\vspace{-5mm}
\label{fig:lifecycle}
\end{figure}

\begin{figure*}[t]
\centering

\begin{minipage}[t]{0.49\textwidth}
\centering
\includegraphics[width=0.7\linewidth]{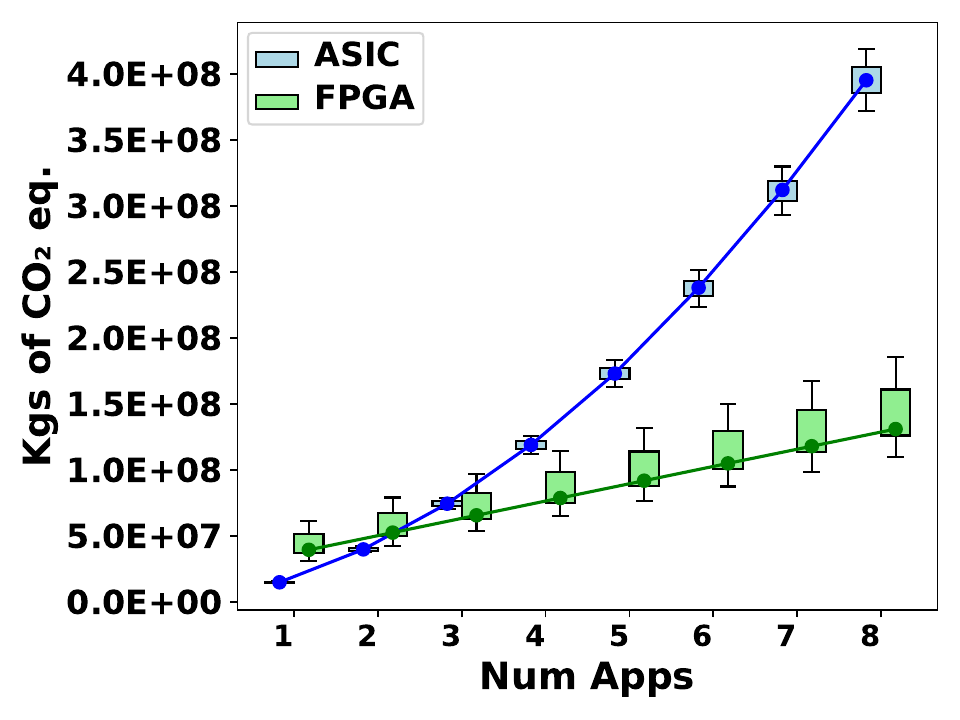}
\caption{Comparing the CFP of FPGAs vs. ASICs at iso-performance while sweeping the number of applications using a probabilistic model (box plot data), and a deterministic model~\cite{greenfpga-dac} (line plot).}
\label{fig:motivate-uncertainities}
\end{minipage}
\hfill
\begin{minipage}[t]{0.49\textwidth}
\centering
\includegraphics[width=\linewidth]{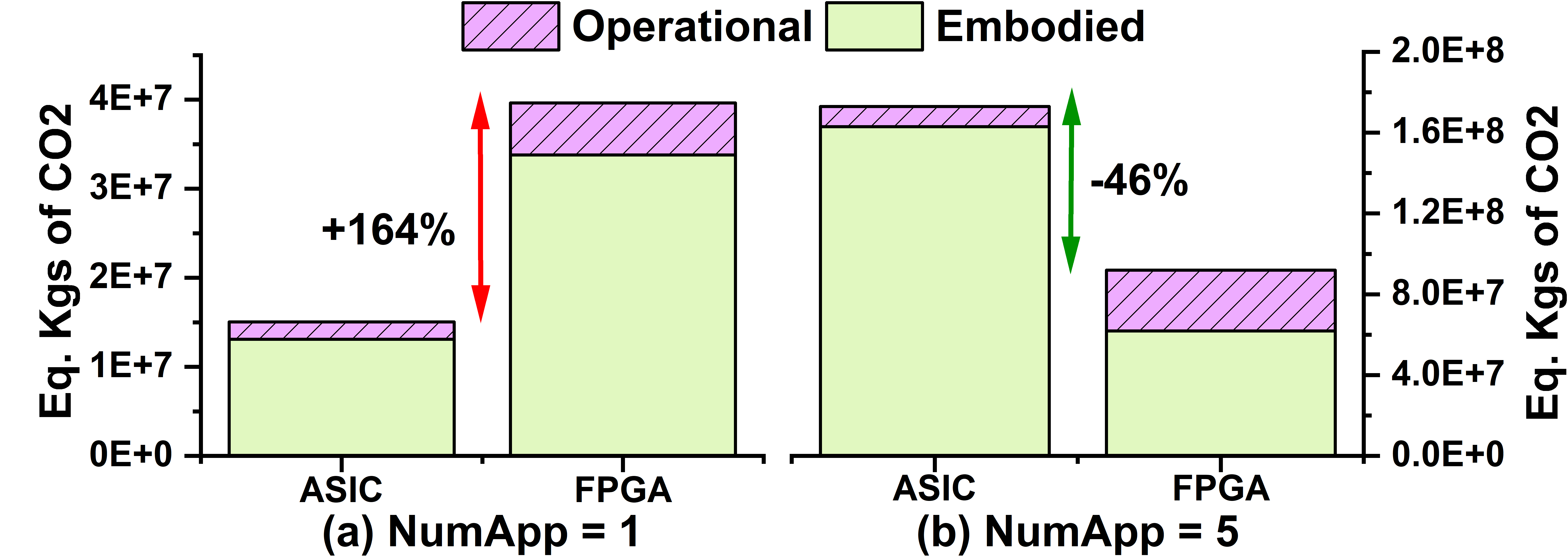}
\caption{CFP comparison between ASIC and FPGA-based computing for a single application and five applications.}
\label{fig:motivation}
\end{minipage}
\vspace{-5mm}
\end{figure*}

Fig.~\ref{fig:lifecycle} shows the complete lifecycle of a computing platform or a chip.  
It outlines the factors contributing to embodied CFP, including design, manufacturing, memory, package, testing, discard, retire, and end-of-life (EOL), and operational CFP tied to end-user activities, reconfigurability, and reliability. Here, retire denotes the terminal handling of the device through discard or recycle, and EOL refers to replacement-related lifecycle effects caused by failure, obsolescence, or upgrade-driven turnover.
The key difference compared to ASICs is that FPGAs, CPUs, and GPUs can be reconfigured or reprogrammed to be used across multiple applications, as shown in the green box, while  ASICs are at the end of life once the application's lifetime is complete. In this paper, we introduce GreenFPGA, a tool to assess the CFP of FPGA-based computing throughout its lifecycle shown in Fig.~\ref{fig:lifecycle}. We focus on evaluating the CFP of FPGAs, and develop new models that can account for these FPGA-related distinctions.
While FPGAs exhibit larger physical size and lower energy efficiency compared to equivalent
ASICs, leading to increased embodied and operational CFP,
they offer compelling sustainability advantages:

\begin{itemize}[nosep, leftmargin=*]
    \item Reconfigurability: FPGAs are reconfigurable and adaptable for multiple applications post-manufacturing, a unique feature absent in ASICs and limited in GPUs.
    \item Extended lifespan: FPGAs typically have longer lifespans and use cases lasting 12 to 15 years~\cite{altera_fpga_lifetime} as they can be reconfigured, compared to ASICs and GPUs that become obsolete with rapidly changing applications (2 to 8 years).
    \item Energy-efficient: FPGAs can offer higher energy-efficiency compared to CPUs and GPUs in some scenarios, as they can be programmed and tailored at the hardware level to meet specific task requirements, minimizing unnecessary operations and energy usage.
\end{itemize}
Over time and across diverse applications, unlike ASICs, the embodied CFP incurred during the manufacturing and design of FPGAs, GPUs, and CPUs can be amortized over their lifespans and across their uses.
Therefore, they make a compelling case from a sustainability perspective for analyzing the CFP. Using GreenFPGA, we analyze scenarios to determine when FPGAs are more sustainable alternatives for computing than ASICs, CPUs, and GPUs. For instance, Fig.~\ref{fig:motivate-uncertainities} compares FPGAs and ASICs under conditions where they have similar performance (iso-performance) for a given application (iso-performance).  The line plot suggests FPGAs are more sustainable for use cases with more than three applications (similar experimental setup as Sec.~\ref{sec:res-asic-v-fpga}).  Similarly, Fig. \ref{fig:motivation} compares ASICs with FPGA at iso-performance~\cite{tian_tan_phd_utexas} (similar experiment setup as Sec.\ref{sec:res-asic-v-fpga}). When the number of applications (NumApp) is equal to 1, FPGAs have higher CFP compared to ASICs (+164\%) as they have larger areas and lower energy efficiencies for these testcases. However, when NumApp = 5, FPGAs exhibit a lower CFP (-46\%) than ASICs due to amortization of the embodied CFP.   We build GreenFPGA, a CFP modeling tool for FPGAs, and compare them against CPUs, ASICs, and GPUs.   The key contributions of our work are as follows.

\begin{enumerate}[noitemsep,topsep=0pt,parsep=0pt,partopsep=0pt]
    \item To the best of our knowledge, our work, GreenFPGA, is the first to model and assess the CFP of FPGA across its entire lifespan by accounting for the unique aspects of FPGA-based computing, including CFP overheads from reconfiguring the FPGA and application development time.

    \item Unlike prior work in CFP modeling of chips~\cite{sudarshan2023ecochip, act, 3dcarbon, greenfpga-dac, refresh-fpga},
    GreenFPGA models replacement effects (EOL of devices) due to device failure, obsolescence, refresh/upgrade, and is the first to incorporate reliability-aware degradation of energy efficiency for operational carbon.
    
    \item We develop a more robust model for carbon modeling of the design phase of a chip's lifecycle compared to~\cite{sudarshan2023ecochip} based on industry (fabless design houses) reports, and also develop a model for testing-related embodied CFP.


    
    \item Given that RAM can contribute significantly up to ~38\%~\cite{google_sustainability_tpu} to the overall total CFP, our framework now includes the embodied CFP of RAM and not just the processor as in prior works~\cite{greenfpga-dac, act}.


    \item We extend  the deterministic model in~\cite{greenfpga-dac} to a probabilistic carbon modeling framework, inspired by~\cite{probabilistic-act, wenkai-operational-uncertainity}, to capture uncertainty in embodied and operational carbon estimation. Our framework now also accounts for FPGA-specific uncertainty sources arising from reconfigurability.
   
    \item We compare ASICs, CPUs, and GPUs with FPGAs, highlighting scenarios where FPGAs offer probabilistically more sustainable alternatives at iso-performance.

\end{enumerate}

\noindent
GreenFPGA is open-source and available to the public~\cite{github_greenfpga}.

\section{Related work}
\label{sec:related-work}

A summary of prior work and the CFP modeling dimensions covered by each framework is provided in Table~\ref{tab:prior_work_comparison}. The table compares existing frameworks across the key components required for a complete lifecycle-aware CFP analysis. The different components of CFP are illustrated in Fig.~\ref{fig:lifecycle}. We highlight the distinguishing features of these models in terms of both embodied and operational CFP, as summarized in Table~\ref{tab:prior_work_comparison}.

ACT~\cite{act, mobile-cfp-survey} developed a data-driven model for embodied CFP estimation using publicly available sustainability reports~\cite{apple-report, TSMC-csr_all, imec-wp}.  ECO-CHIP~\cite{sudarshan2023ecochip,3dcarbon, carbon-path} explores the transition toward chiplet-based architectures and proposes frameworks to estimate both embodied and operational CFP, including the impact of advanced packaging, as reflected in Table~\ref{tab:prior_work_comparison}. ECO-CHIP models memory CFP as an additional chiplet, rather than explicitly capturing memory impact based on CFP per GB. REFRESH-FPGA~\cite{refresh-fpga} proposes a refresh interposer approach that enables the reuse of retired FPGA dies via 2.5D integration, thereby extending operational lifetime and amortizing the embodied CFP over a longer usage period. Additionally,~\cite{cgra-sustainability-iccad-lieven} demonstrates that coarse-grained reconfigurable arrays (CGRAs) can be more sustainable than domain-specific accelerators through relative CFP comparisons but does not explicitly model the absolute CFP, as shown in Table~\ref{tab:prior_work_comparison}, both ~\cite{refresh-fpga,cgra-sustainability-iccad-lieven} do not provide a complete lifecycle CFP formulation. U-DUCT~\cite{wenkai-operational-uncertainity}, along with~\cite{probabilistic-act, carbon-set}, models carbon under uncertainty, capturing spatial, temporal, process-driven, and system-level variability. These works highlight the significant variability that can arise in both embodied and operational CFP, as well as the challenges in validating such models.

Despite these advances, prior work does not provide a unified framework for complete lifecycle CFP analysis of FPGAs. Our prior work~\cite{greenfpga-dac} presents a comprehensive CFP modeling framework that captures embodied CFP, including manufacturing, design, packaging, and retire phases, along with operational carbon. This framework is used to analyze the conditions under which FPGAs are more sustainable computing platforms compared to ASICs.

In this work, we extend~\cite{greenfpga-dac} in several important directions. First, we incorporate device-replacement modeling to capture reasons for the EOL of a device. Second, we introduce uncertainty-aware CFP modeling to account for variability across system and process parameters. Third, we include additional embodied CFP components such as testing CFP and RAM manufacturing CFP. Fourth, we incorporate models for aging-related degradations in the energy efficiency of the device. Finally, we extend the analysis beyond FPGAs and ASICs to include GPUs and CPUs, enabling a broader comparison across computing platforms. These additions, and the expanded coverage of GreenFPGA relative to prior work, are summarized in Table~\ref{tab:prior_work_comparison}.

\begin{table}[ht]
\centering
\caption{Comparison of prior work across CFP modeling dimensions.}
\resizebox{\linewidth}{!}{%
\setlength{\arrayrulewidth}{1pt}
\begin{tabular}{!{\vrule width 2pt}l|l!{\vrule width 2pt}c c c c c c c!{\vrule width 2pt}}
\noalign{\hrule height 2pt}
\textbf{Category} & \textbf{Modeling Aspect} 
& \textbf{ACT~\cite{act}} 
& \textbf{ECO-CHIP~\cite{sudarshan2023ecochip}} 
& \textbf{U-DUCT~\cite{wenkai-operational-uncertainity}} 
& \textbf{REFRESH-FPGA~\cite{refresh-fpga}} 
& \textbf{~\cite{cgra-sustainability-iccad-lieven}}
& \textbf{~\cite{greenfpga-dac}} 
& \textbf{GreenFPGA (This work)} \\
\noalign{\hrule height 2pt}

\multirow{9}{*}{\shortstack{Embodied}}
& Manufacturing CFP                   & \ding{51} & \ding{51} & \ding{51} & \ding{55} & \ding{55} & \ding{51} & \ding{51} \\
& Design CFP                          & \ding{55} & \ding{51} & \ding{55} & \ding{55} & \ding{55} & \ding{51} & \ding{51} \\
& RAM CFP                          & \ding{51} & \ding{51} & \ding{55} & \ding{55} & \ding{55} & \ding{55} & \ding{51} \\
& Package CFP                         & \ding{51} & \ding{51} & \ding{55} & \ding{55} & \ding{55} & \ding{51} & \ding{51} \\
& Testing CFP                         & \ding{55} & \ding{55} & \ding{55} & \ding{55} & \ding{55} & \ding{55} & \ding{51} \\
& Retire CFP                          & \ding{55} & \ding{55} & \ding{55} & \ding{51} & \ding{55} & \ding{51} & \ding{51} \\
& End-of-life CFP                     & \ding{55} & \ding{55} & \ding{55} & \ding{55} & \ding{55} & \ding{55} & \ding{51} \\
& FPGA reconfigurability modeling     & \ding{55} & \ding{55} & \ding{55} & \ding{51} & \ding{51} & \ding{51} & \ding{51} \\
& CPU/GPU programmability modeling  & \ding{55} & \ding{55} & \ding{55} & \ding{55} & \ding{55} & \ding{55} & \ding{51} \\
\hline

\multirow{1}{*}{\shortstack{Embodied+Operational}}
& Probabilistic CFP                   & \ding{55} & \ding{55} & \ding{51} & \ding{55} & \ding{55} & \ding{55} & \ding{51} \\
\hline

\multirow{3}{*}{\shortstack{Operational}}
& Baseline operational CFP            & \ding{55} & \ding{51} & \ding{51} & \ding{55} & \ding{55} & \ding{51} & \ding{51} \\
& Reconfiguration CFP                 & \ding{55} & \ding{55} & \ding{55} & \ding{55} & \ding{55} & \ding{51} & \ding{51} \\
& Reliability-aware CFP               & \ding{55} & \ding{55} & \ding{55} & \ding{55} & \ding{55} & \ding{55} & \ding{51} \\
\noalign{\hrule height 2pt}

\multirow{1}{*}{\shortstack{Framework}}
& Open-source availability             & \ding{51} & \ding{51} & \ding{55} & \ding{55} & \ding{55} & \ding{51} & \ding{51} \\
\hline

\hline
\end{tabular}
}
\label{tab:prior_work_comparison}
\end{table}

\section{GreenFPGA: Models for CFP assessment}
\label{sec:greenfpga-framework}
\begin{figure}[t]
\centering
\includegraphics[width=0.75\linewidth]{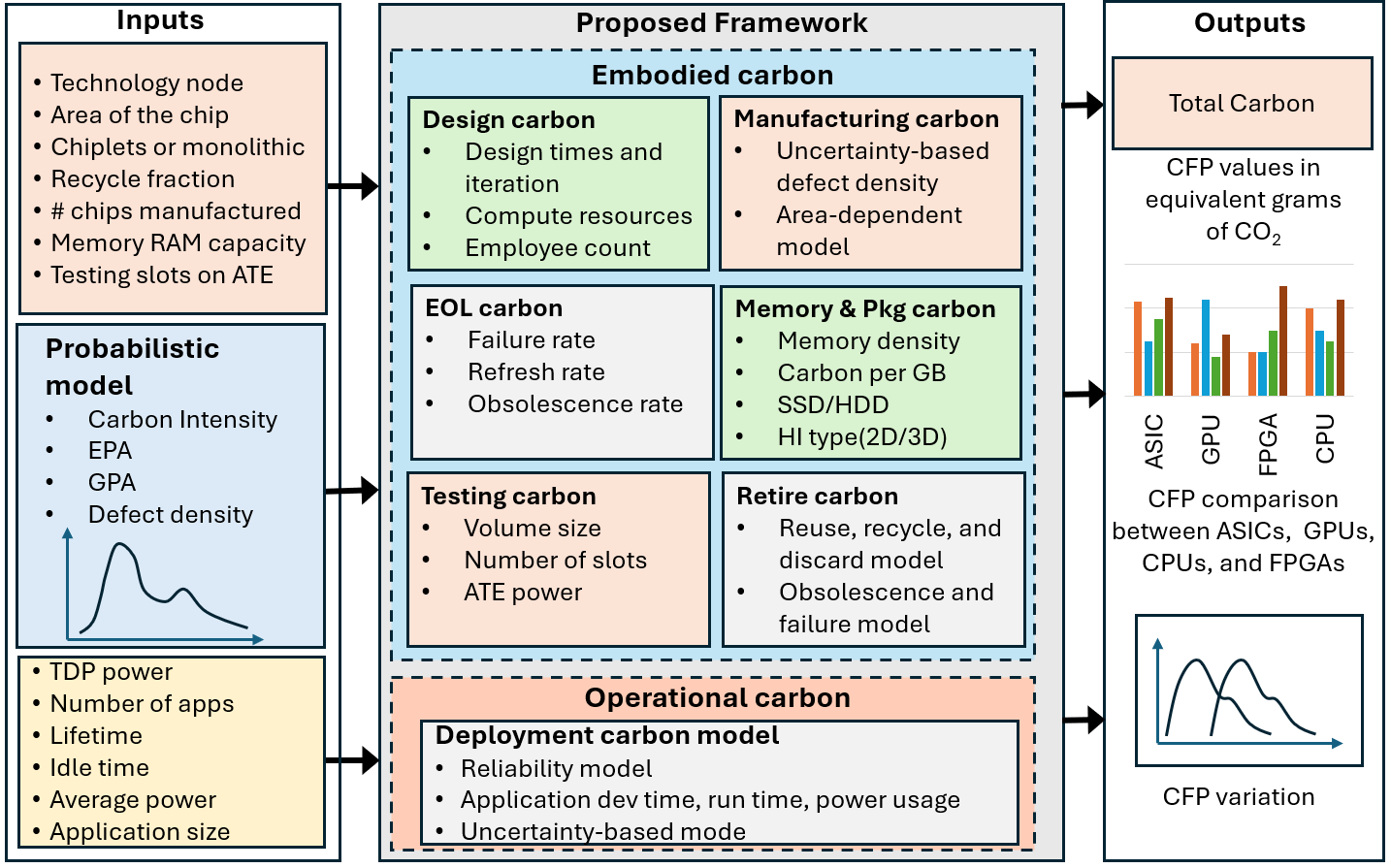}
\caption{GreenFPGA framework, showing the required inputs, probabilistic modeling components, embodied and operational carbon models, and the resulting CFP outputs for comparative platform analysis.}
\vspace{-5mm}
\label{fig:top-level}
\end{figure}

A high-level overview of GreenFPGA is shown in Fig.~\ref{fig:top-level}. As illustrated in the input panel of Fig.~\ref{fig:top-level}, the framework takes device and architecture parameters such as technology node, chip area, packaging style, recycle fraction, number of chips manufactured, RAM capacity, and testing-related parameters. GreenFPGA also incorporates a probabilistic input model for uncertain parameters, including carbon intensity, energy consumed per unit area (EPA), gas emission per unit area (GPA), and defect density. These parameter distributions are generated either using kernel density estimation (KDE) when sufficient data are available or by inferring suitable prior distributions based on the uncertainty model chosen for each parameter. In addition, deployment-related inputs such as TDP power, number of applications, operational lifetime, idle time, average power, and application size are used to evaluate operational CFP.

The model structure of GreenFPGA is summarized in the center panel of Fig.~\ref{fig:top-level}. The framework is divided into embodied carbon and operational carbon components. The embodied carbon model includes design, manufacturing, RAM memory and packaging, testing, retire, and EOL carbon models. In particular, GreenFPGA introduces a more detailed design-phase CFP model based on industrial reports from design houses, incorporates manufacturing and packaging models from prior work~\cite{act, imec-dtco,sudarshan2023ecochip}, includes memory-related CFP modeling based on~\cite{dirty-ssd-carbon,act,mobile-cfp-survey,mem-ram-udit1}, and adds a testing model that accounts for automated test equipment (ATE) usage and test time. As also indicated in Fig.~\ref{fig:top-level}, the framework includes EOL modeling through failure, obsolescence, and refresh effects, along with retire modeling based on recycle and discard behavior. GreenFPGA adopts the probabilistic model from~\cite{probabilistic-act, wenkai-operational-uncertainity} for uncertainties in embodied CFP models and operational CFP. We further extend uncertainty formulations to FPGA-specific parameters and incorporate them into our analysis. On the operational side, Fig.~\ref{fig:top-level} shows a deployment carbon model that captures application development, runtime usage, reconfigurability, and reliability-aware degradation. The detailed formulations of each of these models are presented in the remainder of this section.

The outputs of the framework are shown in the right panel of Fig.~\ref{fig:top-level}. GreenFPGA produces total carbon estimates that support comparative CFP analysis across ASIC, GPU, CPU, and FPGA platforms at iso-performance, and captures CFP variation under uncertainty. Thus, the outputs in Fig.~\ref{fig:top-level} enable not only absolute CFP estimation for a given design point, but also comparative analysis to identify sustainable deployment regions across computing platforms.

\subsection{GreenFPGA: Total CFP model} 
Given that both embodied CFP and operational CFP contribute to the total CFP.
The total CFP of ASICs to perform $N_\text{app}$ different applications is~\cite{sudarshan2023ecochip}:

\begin{equation}
     C_\text{ASIC} = \sum^{i=N_\text{app}}_{i=1}  (C_\text{emb, i} +    C_\text{deploy, i}) 
    \label{eq:tot-cfp-asic}
\end{equation}

\noindent
where the embodied CFP, $C_\text{emb, i}$ is given by the sum of the CFP from design, manufacturing, and packaging, for application $i$, 
and $C_\text{deploy, i}$ is the CFP from reconfiguration and operation of the ASIC for application $i$ in the field. The models for design, manufacturing, and packaging are available in~\cite{sudarshan2023ecochip}. However, these models do not directly apply to CPUs, GPUs, and FPGAs, which are particularly interesting to model for CFP due to their reconfigurability.

\noindent
We model the total CFP of programmable processors (CPU, GPU, and FPGA) as the sum of the embodied and operational CFP as shown below. However, unlike ASICs (Eq.~\eqref{eq:tot-cfp-asic}), the same FPGA, CPU, or GPU can be reused for different applications, and therefore, the total CFP for $N_\text{app}$ applications using FPGAs is given by:

\begin{equation}
    C_\text{processor} =   C_\text{emb} + \sum^{i=N_\text{app}}_{i=1}    C_\text{deploy, i} 
    \label{eq:tot-cfp-fpga}
\end{equation}

\noindent 
where $C_\text{emb}$ is the embodied CFP, and $C_\text{deploy, i}$ is the CFP from reconfiguration and operation for application $i$, for the FPGA, CPU, or GPU. 
In the rest of this paper, we use the following terminologies, \(T_i\) denotes the lifetime of application \(i\), i.e., the duration for which a single application is deployed and executed on the platform. Since chips are not utilized 100\% of the time over their lifetime, we use \(f_{\text{use}}\) to denote the use time fraction, which captures the fraction of the operational lifetime during which the chip is actively in operation. Throughout this paper, we use these terms to distinguish application-level and device-level time scales in the CFP analysis. \(T_{\text{life}}\) denotes the operational lifetime of the device itself. These terms are used later in the paper in the device-level lifecycle models. In the rest of this section, we detail the models for each of the components in Eq.~\eqref{eq:tot-cfp-fpga}.


\subsection{GreenFPGA: Embodied CFP model}
With embodied CFP dominating the total CFP, particularly in battery-operated devices, and devices on the edge~\cite{chasing-carbon,sudarshan2023ecochip,mobile-cfp-survey}, it is crucial to have models that account for CFP from all activities related to manufacturing, design, retire and packaging. 
In addition, devices may require replacement during the operational lifetime for three main reasons: unexpected hardware failures, technological obsolescence that renders the device insufficient for evolving workload or system requirements, and planned refresh or upgrade cycles introduced to improve capability and efficiency. These reasons determine the lifetime of the device. Therefore, it is important to account for the replacement-driven EOL embodied CFP associated with these lifecycle events.
The embodied CFP for the ASIC, FPGA, GPU, or CPU, $C_\text{emb}$, for an application volume of $N_\text{vol}$ systems is given by: 
\begin{equation}
\label{eq:embodied}
    C_\text{emb}= \\ 
C_{\text{des}} +  (N_\text{vol}+N_\text{EOL}) C_\text{device} 
\end{equation}
\noindent
where $C_{\text{des}}$ is the design CFP and is a one time cost that is amortized across all devices manufactured, and $C_{\text{device}}$ is the per-device embodied CFP contribution (Eq.~\ref{eq:per-dev}), $N_\text{vol}$ is the total volume manufactured initially for a specific application, and $N_\text{EOL}$ is the \textbf{\textit{expected}} number of device replacements due to EOL over the operational lifetime and is modeled by:
\begin{equation}
\label{eq:Neol}
    N_\text{EOL} = \sum_{1}^{N_\text{vol}} T_{\text{life}}\,\lambda_{\text{EOL}}
\end{equation}
\noindent
where $T_{\text{life}}$ is the operational lifetime of the device, measured in years (yr), and $\lambda_{\text{EOL}}$ is the total end-of-life replacement rate per unit time, measured in $\text{yr}^\text{-1}$, and is defined by:
\begin{equation}
\label{eq:lam-tot}
\lambda_{\text{EOL}} = \lambda_{\text{fail}} + \lambda_{\text{obsol}} + \lambda_{\text{upgrade}}
\end{equation}

\noindent
where $\lambda_{\text{fail}}$ is the hardware failure rate per unit time~\cite{failure-1,failure-2}, $\lambda_{\text{obsol}}$ is the obsolescence rate per unit time~\cite{obsolete-1,obsolete-2}, and $\lambda_{\text{upgrade}}$ is the rate associated with planned technology upgrades per unit time~\cite{refresh-1,refresh-2,refresh-3}. We assume a stochastic distribution for these parameters and accounting for $\lambda_\text{EOL}$ is important because replacement events can significantly increase the embodied CFP over the operational lifetime of the device, especially in long-lived and large-volume systems. Therefore, a longer operational lifetime ($T_\text{life}$) is not necessarily always beneficial from a CFP perspective, since for sufficiently large $\lambda_\text{EOL}$, the expected number of replacements also increases, leading to higher embodied carbon. The per-device embodied CFP ($C_\text{dev}$) is given by:
\begin{equation}
\label{eq:per-dev}
    C_\text{device}= 
 N_\text{proc} \left ( C_{\text{mfg}} + C_\text{pkg} + C_\text{retire} + C_\text{test} + C_\text{mem}\right )
\end{equation}
where $C_\text{mfg}$ is the manufacturing CFP that accounts for activities related to fabrication,  and $C_{\text{pkg}}$ is the CFP from package manufacture and assembly of the FPGA, GPU, CPU, or ASIC, $C_\text{retire}$ is the retire CFP to model recycle and discard activities,  $C_\text{test}$ is the testing CFP per chip, and $C_\text{mem}$ is the CFP contribution of the off-chip RAM associated with the device. For certain applications, iso-performance comparisons between the processing platforms
require more than one FPGA, as the ASIC counterparts are either at reticle limits or have extremely high performance. Therefore,  we define $N_\text{proc}$ as the number of devices of same type required for a given application for iso-performance and is given by $  \left\lceil\frac{\text{app}_\text{size}}{\text{cores}_\text{capacity}}  \right\rceil$ where the application size and cores  
capacity are specified in terms of equivalent logic gates. 



\begin{figure*}[t]
\centering
\begin{minipage}[t]{0.4\textwidth}
    \centering
    \includegraphics[width=\linewidth]{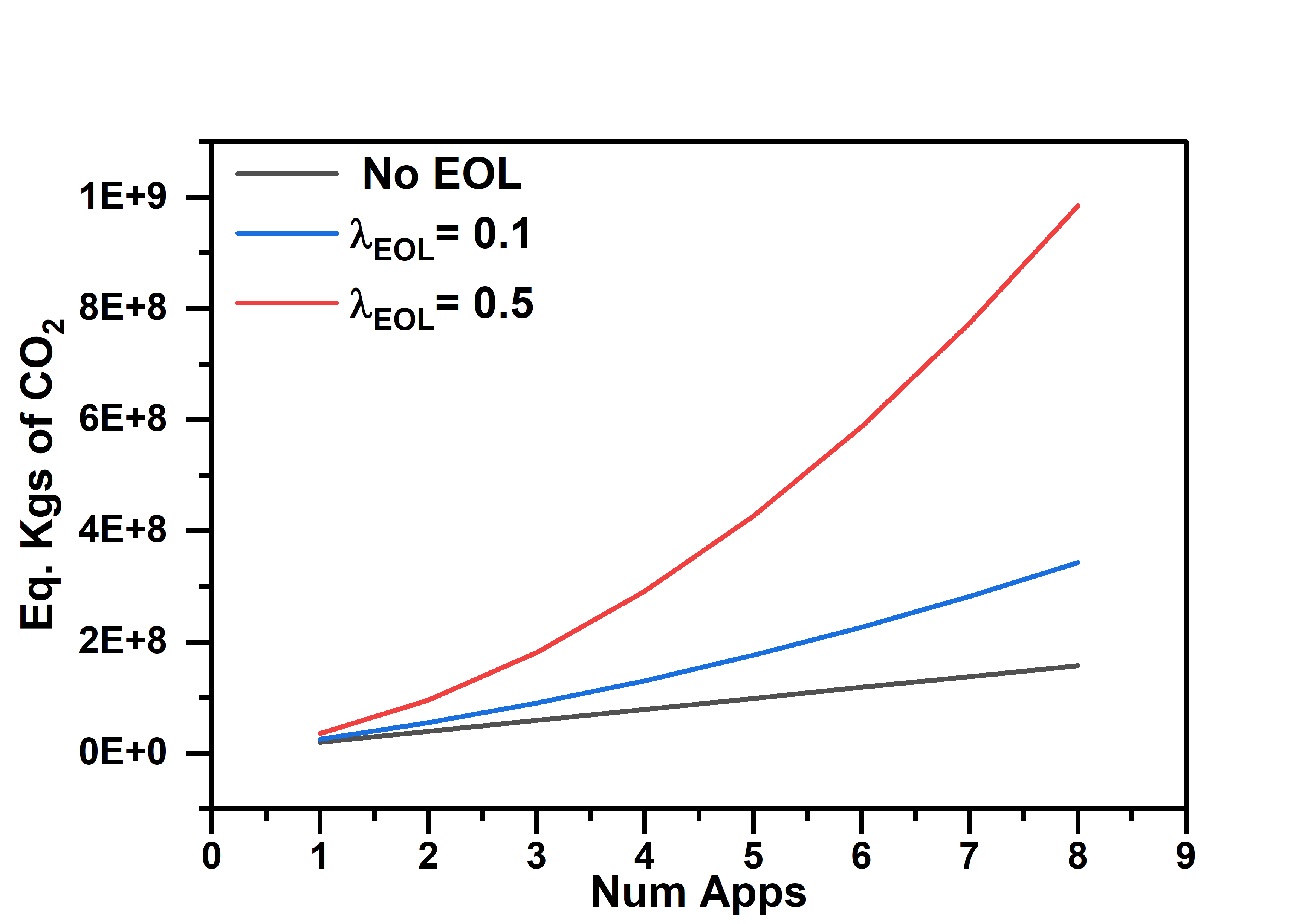}
    \vspace{-7mm}
    \captionof{figure}{Variation in total CFP with number of applications for different $\lambda_\text{EOL}$ rates.}
    \label{fig:moti-eol-rel}
\end{minipage}
\hfill
\begin{minipage}[t]{0.52\textwidth}
    \centering
    \includegraphics[width=\linewidth]{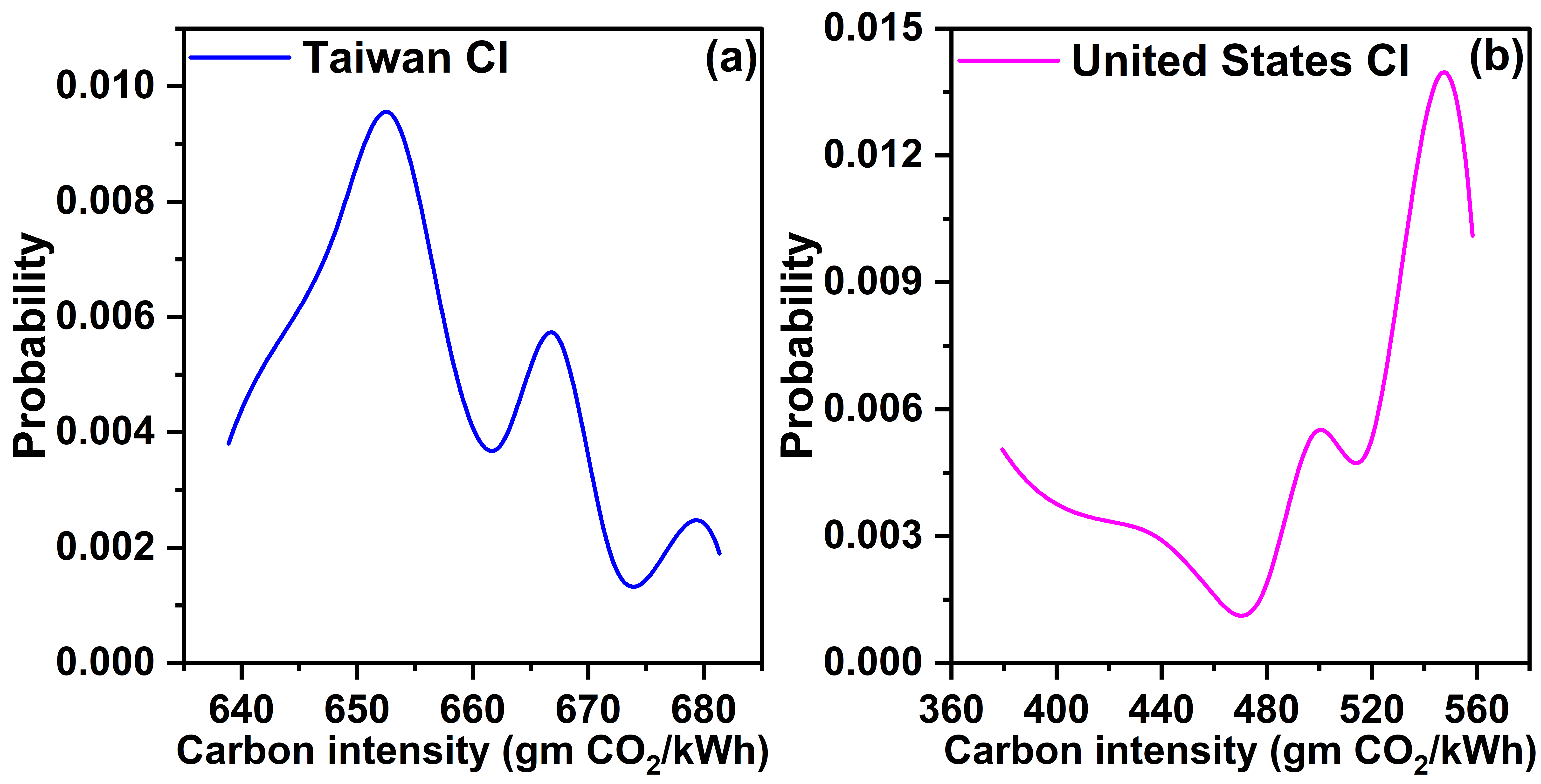}
    \vspace{-7mm}
    \captionof{figure}{KDE of carbon intensity (CI) after considering temporal variations between 2000--2023 for (a) Taiwan and (b) United States~\cite{carbon-intensity-data-source}.}
    \label{fig:kde-carbon-intensity}
\end{minipage}
\vspace{-6mm}
\end{figure*}

Prior work in~\cite{greenfpga-dac} did not account for EOL-driven replacement CFP. Fig.~\ref{fig:moti-eol-rel} highlights the importance of incorporating EOL CFP by showing how the total CFP varies with the number of applications under different end-of-life replacement rates ($\lambda_{\text{EOL}}$). In this experiment we fix application volume to be 1M, set the application lifetime to be 2 years per application, assume an ASIC with area 200 $mm^2$ and power 10 W, and sweep the number of applications from 1 to 8. We compare no-EOL case against $\lambda_{\text{EOL}}$ = 0.1 and $\lambda_{\text{EOL}}$ = 0.5.
As shown in Fig.~\ref{fig:moti-eol-rel}, higher $\lambda_{\text{EOL}}$ leads to substantially higher total CFP because more frequent device turnover introduces additional embodied carbon from replacement devices. The gap between the no-EOL curve and the EOL-aware curves widens as the number of applications increases, indicating that neglecting EOL effects can significantly underestimate lifecycle CFP. This trend is consistent with Eq.~\ref{eq:Neol}: as the effective operational lifetime over which the device is used increases, the expected number of replacements, $N_{\text{EOL}}$ also increases, which in turn increases the embodied CFP contribution. Therefore, for a complete lifecycle assessment, it is important to explicitly model EOL-driven replacement effects in GreenFPGA. This motivates explicit EOL modeling in GreenFPGA. Accordingly, all analyses and results in Sec.~\ref{sec:results} include the impact of EOL-driven replacement effects to support a more accurate lifecycle CFP assessment.

\noindent
{\em (1) Design CFP:} The activities related to chip design include architectural development, RTL, verification, synthesis, simulations, place and route, and various analyses, tests, and post-silicon validation. These activities are performed by large design houses where several engineers work on a common product. While the only existing prior art that models design CFP has used a simplified model that relies on the number of logic gates only~\cite{sudarshan2023ecochip}, the model is difficult to validate. GreenFPGA models the design CFP based on the energy usage of large design houses obtained from sustainability reports~\cite{amd-latest-report, microchip-report, nvidia-report}, the number of products (chips) designed, the fraction of energy coming from renewable resources, the number of employees working on the specific product, and the size of the chip (number of logic gates). The design CFP $C_\text{des}$ for ASIC, FPGA, CPU, or GPU
is given by:

\begin{equation}
    C_\text{des} = C_\text{emp} \times N_\text{emp, des}  \times \frac {N_\text{gates}}{N_\text{gates, proj}}  \times T_\text{proj}
    \label{eq:des}
\end{equation}

\noindent
where $C_\text{emp}$ is the CFP per employee per year, the $N_\text{emp, des}$ is the average number of employees per chip to be designed,  and $N_\text{gates}$ is the number of logic gates in the chip, and $N_\text{gates,  proj}$ is the average number of gates per chip in each project (CPU, GPU, FPGA, and ASIC), and $T_\text{proj}$ is the duration of the chip design project and is given by:

\begin{equation}
    T_\text{proj} = \sum_{i=1}^{N_u} T_{\text{IP},i} + T_{SOC}
    \label{eq:tproj}
\end{equation}

\noindent
where $T_\text{IP}$ is the time taken for the design and verification of a particular Intellectual Property (IP) block, $N_u$ is the total number of unique IPs in the design, $T_{SOC}$ is the time taken for the design and verification of the top-level SoC. Leveraging pre-designed and pre-verified IP can help amortize the design CFP of multiple IPs across different SoCs.

The $C_\text{emp}$ is obtained from industry reports of fabless design houses and is given by $C_\text{emp} = \frac{E_\text{des}}{N_\text{emp,total}} \times C_\text{src, des}$ where $E_\text{des}$ is the total electric energy utilized by the design house per year, $N_\text{emp,total}$ is the total number of employees in the company, and $C_\text{src, des}$ is the carbon intensity for the energy source.  $C_\text{src, des}$ will be lower for renewable resources and higher for non-renewable resources~\cite{chasing-carbon}. Since the carbon intensity varies across time, this is modeled as a KDE as shown in Fig.~\ref{fig:kde-carbon-intensity} for the variations in (a) Taiwan and (b) the United States. These values have been obtained from~\cite{carbon-intensity-data-source}.


Unlike~\cite{sudarshan2023ecochip}, GreenFPGA models the design CFP from industry sustainability reports~\cite{nvidia-report, microchip-report, amd-latest-report}, which is a more reliable source as it utilizes energy values mentioned in the reports and also incorporates uncertainty in carbon intensity.

\noindent
{\em (2) Manufacturing CFP}:
GreenFPGA employs manufacturing CFP models from~\cite{sudarshan2023ecochip, imec-dtco}, where $C_\text{mfg}$ is given by the product of CFP per unit area of manufacturing and the area of the die $\text{CFPA} \times A_\text{die}$. $\text{CFPA}$ is given by:

\begin{align}
    \text{CFPA} & = \frac{(C_{\text{mfg, src}} \times \text{EPA}(p)  + C_{\text{gas}} + C_{\text{materials}})}{Y(d, p)}  
    \label{eq:CFPA}
\end{align}

\begin{figure*}[t]
\centering
\begin{minipage}[t]{0.48\textwidth}
    \centering
    \includegraphics[width=\linewidth]{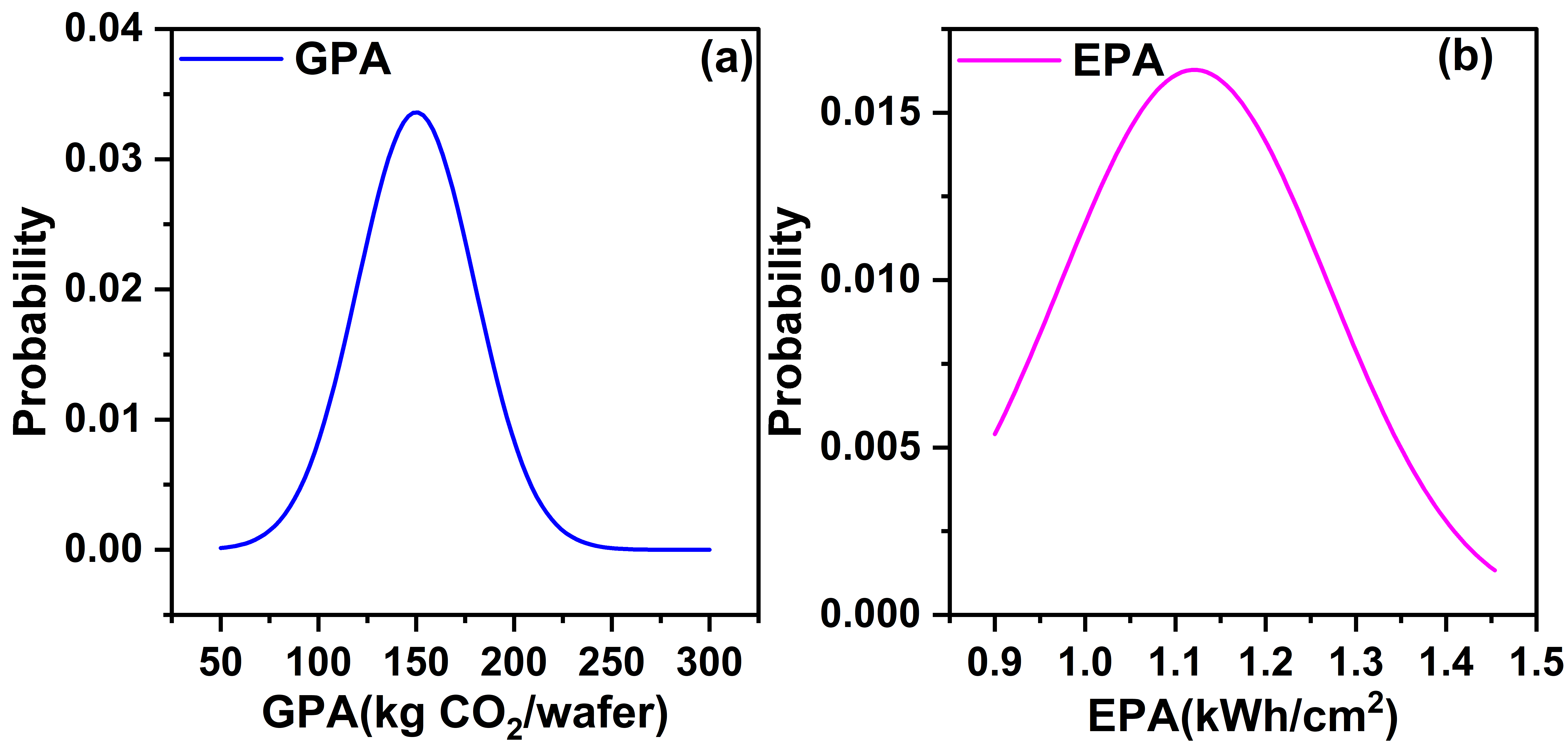}
    \vspace{-6mm}
    \captionof{figure}{Distributions of 10nm GPA and EPA values: (a) normal GPA distribution and (b) KDE for EPA from~\cite{probabilistic-act}.}
    \label{fig:epa-gpa-kde}
\end{minipage}
\hfill
\begin{minipage}[t]{0.48\textwidth}
    \centering
    \includegraphics[width=\linewidth]{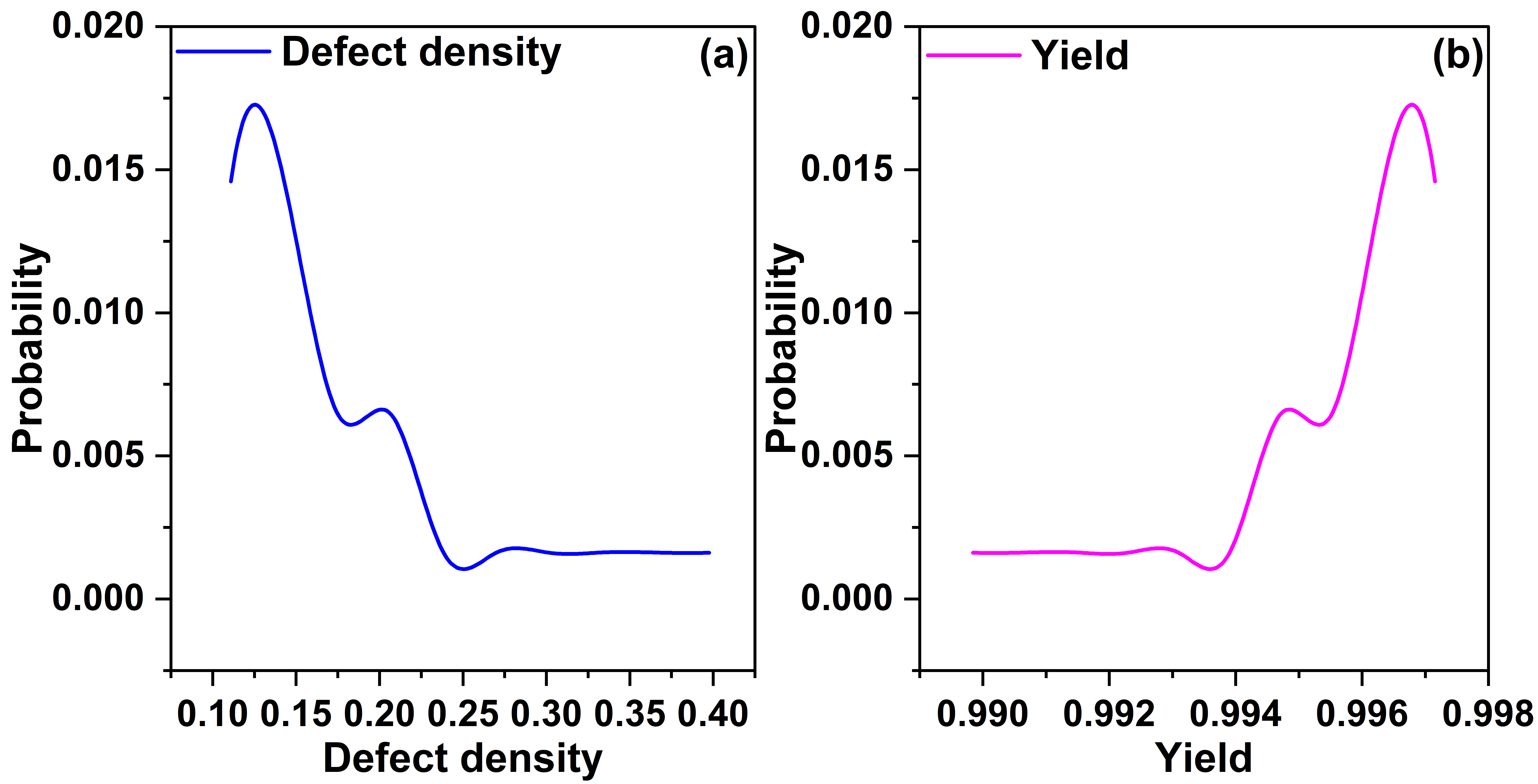}
    \vspace{-6mm}
    \captionof{figure}{(a) KDE of defect densities for the 10nm TSMC technology node~\cite{def-den-tsmc} and (b) yield for the Intel i9 14th Gen processor~\cite{intel-i9,probabilistic-act}.}
    \label{fig:defect-densities-kde}
    \vspace{-8mm}
\end{minipage}
\vspace{-10mm}
\end{figure*}

\noindent
$C_\text{mfg, src}$ is carbon intensity which depends on the energy source of the fab (i.e., renewable or non-renewable), which converts the energy consumed into carbon emissions and is modeled as the KDE shown in Fig.~\ref{fig:kde-carbon-intensity}. $\text{EPA}$ is the energy consumed per unit area during manufacturing of process $p$ and derived from~\cite{imec-dtco,TSMC-csr_all}.
$C_\text{gas}$ is the CFP from the greenhouse gas emissions per unit area (GPA). 
We use a probabilistic model for these parameters to account for uncertainty in manufacturing processes over time and in different regions.  
Similar to~\cite{probabilistic-act}, GreenFPGA models GPA as a Gaussian distribution~\cite{imec-dtco,ipcc-epa-gpa} and EPA using KDEs of temporally distributed data.  
Fig.~\ref{fig:epa-gpa-kde} shows the distributions for GPA and EPA values for 10nm technology node (For GPA we use 300mm wafer size). 
$C_{\text{materials}}$ is the CFP of sourcing the materials for fabricating the chip per unit area. 
$Y(d, p)$ is the yield of the unit area die $d$ in process $p$~\cite{yield-eqn-source}: 

\begin{equation}
\label{eq:yield-eqn}
    Y = \left ( 1 + \frac{A_{d} \times D_0}{\alpha} \right ) ^ {-\alpha} 
\end{equation}
where $A_{{d}}$ is the area of the die, $\alpha$ is the clustering parameter, and $D_0$ is defect density for the technology node, which is also modeled as a KDE as shown in Fig.~\ref{fig:defect-densities-kde}.

%

Further, GreenFPGA models the materials fetched from recycled sourcing and newly extracted materials. We utilize data from~\cite{apple-recycle,recycle-cfp} to extract the percentage of materials that can be recycled ($\rho$) per unit area of product and scale the CFP from sourcing materials as:

\begin{equation}
\label{eq:mfg-tot}
    C_\text{materials} = {\rho} C_\text{materials, recycled} + (1-{\rho})  C_\text{materials, new}
\end{equation}

\noindent
where $C_\text{materials}$ is the component of $C_\text{mfg}$ related to sourcing raw materials, and $C_\text{materials, new}$ is the CFP when the materials are extracted from source, and $C_\text{materials, recycled}$ is the CFP from using recycled materials per unit area. 

\noindent
{\em(3) Package CFP:} We use the monolithic package CFP model from~\cite{sudarshan2023ecochip} and account for uncertainties in the same way as we do for manufacturing CFP.

\noindent
{\em(4) Retire CFP:} This includes CFP from discarding and a CFP credit for recycling a fraction ($\delta$) of chips and is given by:

\begin{equation}
\label{eq:eol}
C_\text{retire} =  (1-\delta ) C_\text{dis} - \delta C_\text{recycle} 
\end{equation}

\noindent
where $\delta C_\text{recycle}$ is the CFP credit from recycling, $(1-\delta )C_\text{dis}$ is the CFP of discarding. We acquire the $C_\text{dis}$ and $C_\text{recycle}$ ranges from government-released public reports~\cite{epa-gov-recycle}, in which the values are reported in CO$_2$/ton and are converted in our analysis to per-device values.

\noindent
{\em (5) Testing CFP:} As chip designs approach reticle limits and their complexity grows, testing becomes increasingly challenging. Automated test equipment (ATE) is widely used to thoroughly test and validate these chips. The test duration for each chip is influenced by factors such as the chip's size and complexity, the number of test vectors required, and the specific type of tests performed on the ATE.

\noindent
Equation~\eqref{eq:ctest2} models the total testing duration where $N_\text{vol}$ is the total volume of chips manufactured, $N_\text{slots}$ is the number of slots on the ATE used, $T_\text{testing,i}$ is the time it takes to test one chip for a particular test type, ${test\_type}$ is the total types of testing performed on the ATE. $T_\text{overhead}$ is the additional overhead involved with loading the ATE and transitioning test types. One can determine the total testing time per chip by dividing the $T_\text{total-test}$ by total chip volume.

\begin{equation}
\label{eq:ctest2}
T_\text{total-test} =  ( \sum_{i=1}^{\text{test\_type}} \frac{N_{vol}}{N_\text{slots}} \times T_\text{testing,i} )  + T_\text{overhead} 
\end{equation}

\noindent
Based on the type of ATE used and test type performed, the overall energy usage ($E_\text{total-test}$) for testing depends on power consumed by the ATE machine ($P_\text{total-test}$), and the total testing time ($T_\text{total-test}$). 

\begin{equation}
\label{eq:cfp-test}
C_\text{test,total} =  C_\text{test,src} \times P_\text{total-test} \times T_\text{total-test}
\end{equation}

\noindent
\(C_{\text{test,total}}\) denotes the total testing CFP across all chips tested on the ATE, and $C_\text{test,src}$ is the carbon intensity of the energy source for the testing equipment and modeled as the KDE in Fig.~\ref{fig:kde-carbon-intensity}. The per-device testing CFP, \(C_{\text{test}}\), is obtained by dividing \(C_{\text{test,total}}\) by the total number of tested units, including both the initial manufactured volume and the EOL replacement volume.

\noindent
{\em (6) Memory CFP:}  In GreenFPGA, we have models that include the embodied CFP of off-chip memory, including DRAM (DDR or HBM). The contribution of memory CFP is both substantial and highly dependent on memory technology and capacity (mem\_capacity). Utilizing CFP per GB (CFPGB) values from~\cite{act,dirty-ssd-carbon,mobile-cfp-survey,mem-ram-udit1}, we find that in Intel i9-based systems~\cite{intel-i9}, the processor CFP is approximately 14.5 kg CO$_2$-eq, while a DDR4 configuration with 128GB memory adds about 7 kg CO$_2$-eq, and a DDR5 configuration with 192GB adds about 10 kg CO$_2$-eq. This corresponds to memory CFP contributions of roughly 50\% to over 70\% of the processor CFP with an experimental setup similar to that defined in Section~\ref{sec:setup}. These observations highlight that memory CFP is not only non-negligible but also highly sensitive to system design choices, making it essential to model memory CFP. We model memory CFP as $C_\text{mem} =  \text{CFPGB} \times \text{mem\_capacity}$.



\subsection{GreenFPGA: Deployment CFP model}
GreenFPGA models the deployment CFP, $C_\text{deploy}$, as the sum of the CFP from field operation (product use) and the reconfiguration application development and is given by: 
\vspace{-2mm}

\begin{equation}
    C_\text{deploy}= N_\text{vol} \times C_\text{op} + C_\text{reconfig}
\end{equation}

\noindent
where $N_\text{vol}$ is the number of chips (ASICs, GPUs, CPUs, or FPGAs) manufactured, $C_\text{op}$ is the operational CFP during use of the chip, and $C_\text{reconfig}$ is the reconfiguration CFP.




\noindent 
(1) \textit{Operational CFP:} The operational CFP, \(C_\text{op}\), captures the carbon footprint associated with energy consumed during device usage over its operational lifetime. It is modeled as the product of the carbon intensity of the energy source during use ($C_\text{src, use}$), and the total energy consumed during operation ($E_\text{use}$):
\begin{equation}
\label{eq:cop_basic}
    C_\text{op} = C_{\text{src,use}} \times\, E_\text{use} 
\end{equation}
where $E_\text{use}$ depends on the device power consumption and the duration of use~\cite{sudarshan2023ecochip}. 
For more accurate estimation of operational CFP compared to prior work, we incorporate the reliability impact of the device by accounting for aging-induced efficiency degradation over time~\cite{reliability-bti-sachin,reliablity-finfet-vidya}. The lifetime energy consumption is modeled as: 


\begin{equation}
E_{\text{use}} = P_{\text{use}}\, f_{\text{use}} \int_{0}^{T_{\text{life}}} \gamma(t)\,dt
\label{eq:euse_gamma}
\end{equation}

where \(P_\text{use}\) is the nominal power during operation and
\(T_\text{life}\) is the operational lifetime. 
We model time-dependent aging effects through $\gamma(t)$ while treating $P_\text{use}$ as a nominal constant. This is because obtaining reliable instantaneous power profiles $P_\text{use}(t)$ is difficult across platforms and workloads, and the effect of usage duration is already captured and analyzed separately in GreenFPGA through the use time fraction $f_\text{use}$, since chips are not utilized 100\% of the time over their lifetime~\cite{fuse-data}.
Here, $\gamma(t)$ is an aging factor derived from bias temperature instability (BTI) and hot carrier injection (HCI) models~\cite{reliability-bti-sachin,reliablity-finfet-vidya}, and is used to capture the degradation in operational efficiency over the lifetime of the device by modeling a change in threshold voltage. As the device ages, wear-out mechanisms such as transistor aging, thermal stress, and interconnect degradation can increase the effective energy required to sustain the same computational workload. Accordingly, $\gamma(t)$ scales the nominal power usage profile to reflect this time-dependent degradation.
For an ideal device with no degradation, \(\gamma(t)=1\), the $C_\text{op}$ reduces to the nominal lifetime operational CFP. Therefore, extending operational lifetime is not always favorable, as reliability-driven degradation can increase the operational CFP accumulated over extended use.

\noindent
{\em(2) Reconfiguration CFP:}
For FPGAs, reconfiguring across multiple applications involves RTL (register transfer level) development or HLS (high-level synthesis) flows and hardware-level simulations. 
CPU application development typically involves nightly and weekly regressions for the application itself and the frameworks used for application deployment.
GPUs have extensive feature-rich firmware, and application development includes verifying drivers and kernel libraries; all these add up to a longer regression time compared to CPUs.
In contrast, ASICs utilize software flows with extensive regression testing, as seen in the Google TPU \cite{tpuv4}. 
These different approaches lead to distinct CFPs, because software development is faster than hardware development. We consider this overhead when assessing FPGAs as sustainable computing solutions, as reconfiguration represents a recurring cost per application in Eq.~\eqref{eq:tot-cfp-fpga}. We model $C_\text{reconfig}$ as the product of the power dissipated by the CPU systems used in the reconfiguration development and the carbon intensity of the energy source and the reconfiguration development time, $T_\text{reconfig}$:

\begin{equation}
\label{eq:new-appdev-cfp}
    T_\text{reconfig} = N_\text{app} \times ( T_\text{sw-dev}  + T_\text{compile} + T_\text{reg} ) + N_\text{vol} \times T_\text{app,config}  
\end{equation}

\noindent
where $N_\text{app}$ is the total number of applications.
$T_\text{sw-dev}$ is the time it takes to write an application for a given processor type (FPGA, CPU, or GPU). For example, for FPGAs, this involves writing RTL and performing verification, which is done once per application, as well as synthesizing, placing, and routing, which is performed once per FPGA architecture.
$T_\text{compile}$ is the time it takes to convert the application program written in a high-level programming language to low-level machine code (i.e., from Verilog to bitstream for FPGAs or from CUDA to binary for GPUs).
$T_\text{reg}$ is the time taken for regressions involving hardware simulations, software testing, and firmware verification.
$N_\text{vol}$ is the volume of chips manufactured and $T_\text{app-config}$ is the time it takes to configure or program the FPGA, CPU, or GPU that is deployed.

\section{Experimental setup and testcases}
\label{sec:setup}

\begin{table*}[t]
\centering
\renewcommand{\arraystretch}{1.15}

\begin{minipage}[t]{0.49\textwidth}
\centering
\caption{Input parameter ranges to GreenFPGA.}
\label{tbl:params}
\resizebox{\linewidth}{!}{%
\begin{tabular}{c|crll}
\hline
Model   & Parameter & Value & Unit & Source \\ \hline

\multirow{1}{*}{$C_\text{materials}$}
& $\rho$ & 0 -- 1 & & \cite{apple-recycle}/user-defined \\
\hline

\multirow{1}{*}{$C_\text{gas}$}
& $GPA$ & 50 -- 300 & kg CO$_2$/wafer & \cite{imec-dtco,probabilistic-act} \\
\hline

\multirow{2}{*}{$C_\text{mfg}$}
& $EPA$ & 0.9 -- 1.45 & kWh/cm$_2$ & \cite{ipcc-epa-gpa} \\
& $D_0$ & 0.1 -- 0.4 & def/cm$_2$ & \cite{def-den-tsmc} \\
\hline

\multirow{3}{*}{$C_\text{retire}$}
& $\delta$ & 0.8 & & \cite{epa-gov-recycle}/user-defined \\
& $C_\text{recycle}$ & 7.65 -- 29.83 & $\text{MTCO}_\text{2}\text{E/ton}$ & \cite{epa-gov-recycle} \\
& $C_\text{dis}$ & 0.03 -- 2.08 & $\text{MTCO}_\text{2}\text{E/ton}$ & \cite{epa-gov-recycle} \\
\hline

\multirow{3}{*}{$C_\text{reconfig}$}
& $T_\text{sw-dev}$ & 0.5 -- 2.5 & months & user-defined \\
& $T_\text{compile}$ & 0.5 -- 1.5 & months & user-defined \\
& $T_\text{reg}$ & 0.1 -- 1.5 & months & user-defined \\
\hline

\multirow{3}{*}{$C_\text{des}$}
& $E_\text{des}$ & 2 -- 7.3 & GWh & \cite{amd-latest-report,nvidia-report,microchip-report} \\
& $N_\text{emp, total}$ & 20K -- 160K & employees & \cite{microchip-report,nvidia-report,amd-latest-report} \\
& $T_\text{proj}$ & 1 -- 3 & years & \cite{nvidia-roadmap} \\
\hline

\multirow{1}{*}{$C_\text{mem}$}
& CFPGB & 20 -- 600 & g CO$_2$/GB & \cite{dirty-ssd-carbon,act,mobile-cfp-survey,mem-ram-udit1} \\
\hline

\multirow{2}{*}{$C_\text{op}$}
& $C_\text{src}$ & 30 -- 700 & g CO$_2$/kWh & \cite{act,imec-dtco,carbon-intensity-data-source} \\
& $\gamma(t)$ & $\geq 1$ & & \cite{reliability-bti-sachin,reliablity-finfet-vidya} \\
\hline

\multirow{1}{*}{$C_\text{EOL}$}
& $\lambda_\text{EOL}$ & 0--2 & replacement/yr & \cite{failure-1,failure-2,obsolete-1,obsolete-2,refresh-1,refresh-2,refresh-3}/user-defined \\
\hline

\end{tabular}%
}
\end{minipage}
\hfill
\begin{minipage}[t]{0.49\textwidth}
\centering
\caption{FPGA normalized power and area values for different test cases for ASIC, GPU, and CPU at iso-performance.}
\label{tbl:fpga_norm_closed}
\resizebox{\linewidth}{!}{%
\begin{tabular}{|l|c|c|c|}
\hline
\multicolumn{4}{|c|}{\textbf{ASIC Testcases (Area and Power normalized to ASIC)}} \\ \hline
\textbf{Testcases} & \textbf{DNN~\cite{tian_tan_phd_utexas}} & \textbf{ImgProc~\cite{tian_tan_phd_utexas}} & \textbf{Crypto~\cite{tian_tan_phd_utexas}} \\ \hline
\textbf{Area} & 4 & 7.42 & 1 \\ \hline
\textbf{Power} & 3 & 1.25 & 1 \\ \hline

\multicolumn{4}{|c|}{\textbf{GPU Testcases (Area and Power normalized to GPU)}} \\ \hline
\textbf{Testcases} & \textbf{High Perf-RNN~\cite{eriko-fpga-asic}} & \textbf{Energy Eff-RNN~\cite{eriko-fpga-asic}} & \textbf{LHCb~\cite{gpu-fpga-t3}} \\ \hline
\textbf{Area} & 4.39 & 3.17 & 0.377 \\ \hline
\textbf{Power} & 1.02 & 4.16 & 0.489 \\ \hline

\multicolumn{4}{|c|}{\textbf{CPU Testcases (Area and Power normalized to CPU)}} \\ \hline
\textbf{Testcases} & \textbf{Random Num.\ Gen.~\cite{cpu-fpga-david}} & \textbf{Llama 2~\cite{cpu-fpga-t2}} & \textbf{FIR Filter~\cite{cpu-fpga-t3}} \\ \hline
\textbf{Area} & 2.8 & 0.78 & 2.38 \\ \hline
\textbf{Power} & 0.005 & 0.086 & 0.011 \\ \hline
\end{tabular}%
}
\end{minipage}
\vspace{-5mm}
\end{table*}


\subsection{Setup}
\noindent
GreenFPGA uses several input parameters for its models as listed in Table~\ref{tbl:params} with their corresponding sources.  The input parameters in the case of $\text{EPA(p)}$, gases emitted per unit are manufacturing $C_\text{gas}$ (GPA),  carbon intensity, and defect densities are all distributions with data from~\cite{probabilistic-act} for the 10nm technology node. Certain parameters, such as the reconfiguration time, are difficult to find in the public domain. We assume values for these based on industry input. However, these can be tuned by the user as input parameters.   We used parameters from~\cite{dirty-ssd-carbon, mobile-cfp-survey, act} for memory CFP and assumed all testcases in a 10nm technology node. While some of the input parameters are distributions as specified in Section~\ref{sec:greenfpga-framework}, we only show the \textbf{\textit{expected}} values of CFP due to space constraints. We show the entire distribution of the CFP for one of our experiments and test cases in Section~\ref{sec:probabilistic-model-results}, but for the rest of the experiments, we use the expected value to make our analysis. 

The testcases used in this work correspond to iso-performance normalized area and power ratios between FPGA and the other computing platforms, as summarized in Table~\ref{tbl:fpga_norm_closed}. These ratios are themselves input parameters to GreenFPGA and can be swept more generally whenever iso-performance relative values are known, making the presented testcases illustrative examples of a broader relative-analysis capability. Our analysis is structured as a set of pairwise comparisons, evaluating FPGA sustainability relative to ASICs, GPUs, and CPUs separately, rather than performing a single comparison across all four platforms simultaneously. 

\subsection{Testcases: FPGA vs ASIC}
We compare FPGAs and ASICs at iso-performance, employing power, and area values across three testcases: deep neural networks (DNN), image processing (ImgProc), and cryptography (Crypto) from~\cite{tian_tan_phd_utexas}, considering a 10nm technology node. ~\cite{tian_tan_phd_utexas} emphasizes that while ASICs are designed for flexibility and programmability at the architectural level, they lack reconfigurability at the circuit level post-manufacturing. In contrast, FPGAs offer circuit-level reconfigurability, allowing the architecture of an application to be finely tuned to specific requirements. This insight results in practical area and power metrics ratios between FPGAs and ASICs for the same performance, as highlighted in Table~\ref{tbl:fpga_norm_closed}. The assumption is that FPGAs can adapt to changing application characteristics by loading new configurations, whereas a new ASIC is required for each application change.

\subsection{Testcases: FPGA vs GPU}
\noindent
The work in~\cite{eriko-fpga-asic} compares FPGAs with GPUs, commenting on the performance of AI workloads for high-performance and energy-efficient deployments. For the DNN/AI workloads considered in~\cite{eriko-fpga-asic}, FPGAs are competitive compared to GPUs, offering lower power consumption for similar on-chip memory. For the large hadron collider (LHCb)~\cite{gpu-fpga-t3} testcase, deploying parts of the trigger and pattern-recognition workload on FPGA boards closer to the detector may therefore improve both cost and energy efficiency compared to GPU. GPUs are typically underutilized relative to FPGAs for the considered workloads.
Given the fact that GPUs are more versatile compared to ASICs, in our analysis, we look at cases where GPUs are reused across different applications; more details are explained in Section~\ref{sec:cfp-fpga-gpu}. Table~\ref{tbl:fpga_norm_closed} shows the normalized area and power for iso-performance between FPGA and GPUs for these testcases. To ensure a fair comparison, we use hardware utilization values for area, which indicate the percentage of the overall hardware used to run the application and collect performance data. We scale power and area based on performance and utilization, respectively, for an iso-performance comparison.

\subsection{Testcases: FPGA vs CPU}
Section~\ref{sec:cfp-fpga-cpu} compares CPUs with FPGAs for the random number generation ~\cite{cpu-fpga-david}, Llama 2~\cite{cpu-fpga-t2}, and FIR filter~\cite{cpu-fpga-t3} testcases. Similar to the previous setup, we compare the CPU with the FPGA at iso-performance using performance values from~\cite{cpu-fpga-david, cpu-fpga-t2,cpu-fpga-t3}. The normalized values used for the testcase are in Table~\ref{tbl:fpga_norm_closed}.

\subsection{Industry testcases}
We evaluate the CFP distributions of industry chips listed in Table~\ref{tbl:industry-testcases} with power (TDP) and area values. It includes IndustryASIC based on the Google TPUv4 ASIC~\cite{tpuv4}, IndustryGPU based on NVIDIA H100 GPU~\cite{nvidia-h100}, IndustryCPU based on Intel Core i9-14900KF CPU~\cite{intel-i9}, and IndustryFPGA based on Intel Agilex 7 FPGA I-Series R31B~\cite{agilex7}. For each device, we collect area, power, and technology node information, as shown in Table~\ref{tbl:industry-testcases}, from the available literature.

\begin{table}[t]
  \centering
  \caption{Industry processor summary~\cite{intel-i9, nvidia-h100, agilex7, tpuv4}.}
  \resizebox{0.7\linewidth}{!}{%
  \begin{tabular}{|c|c|c|c|c|}
    \hline
    \textbf{Testcases} & \textbf{IndustryASIC} & \textbf{IndustryGPU} & \textbf{IndustryFPGA} & \textbf{IndustryCPU} \\
    \hline
    Area      & 600 $mm^2$ & 814 $mm^2$ & 550 $mm^2$ & 257 $mm^2$ \\
    \hline
    Power     & 190 W 	& 350 W  & 220 W  & 125 W \\
    \hline
    Tech. Node & 7 nm 	& 5 nm   & 10 nm  & 10 nm \\
    \hline
  \end{tabular}
  }
  \label{tbl:industry-testcases}
\end{table}


\section{Evaluation of GreenFPGA}
\label{sec:results}

\subsection {Comparing CFP of FPGAs and ASICs}
\label{sec:res-asic-v-fpga}
As noted earlier, the embodied CFP and the deployment CFP of an FPGA are higher than that of an iso-performance ASIC because the FPGA required has a larger area and consumes more power.
However,  FPGA reconfigurability can help amortize the embodied CFP over the operational lifetime of the chip, thereby reducing the overall CFP.

We observe the impact of the number of applications (Num Apps) $N_\text{app}$, volume of applications (App Volume)  $N_\text{vol}$, lifetime of each application (App Lifetime) $T_\text{i}$, and the use time fraction $f_\text{use}$ of the device on the  CFP of FPGAs and ASICs.
For each experiment, we define the point at which the CFP of FPGA becomes lower compared to ASIC as the A2F crossover point, and the point at which the  CFP of FPGA becomes higher compared to the ASIC as the F2A crossover point. To understand the impact of each variable, we set up four experiments (A -- D) where the variable of interest is varied while keeping the other three variables constant. We use the following constant values $T_i = 2$ years, $N_\text{vol} = 1e6$, $f_\text{use} = 0.2$, and $N_\text{app} = 5$. 
We use $\lambda_\text{EOL}$ for ASIC to be 0.3 $\text{yr}^\text{-1}$ (3.3 years per replacement) and FPGA to be 0.13$\text{yr}^\text{-1}$(7.5 years per replacement)~\cite{greenfpga-dac,altera_fpga_lifetime,obsolete-2,failure-1,refresh-3}. These parameters are configurable inputs to GreenFPGA and can be adjusted to evaluate different scenarios.

\begin{figure}[t]
\centering
\includegraphics[width=0.6\linewidth]{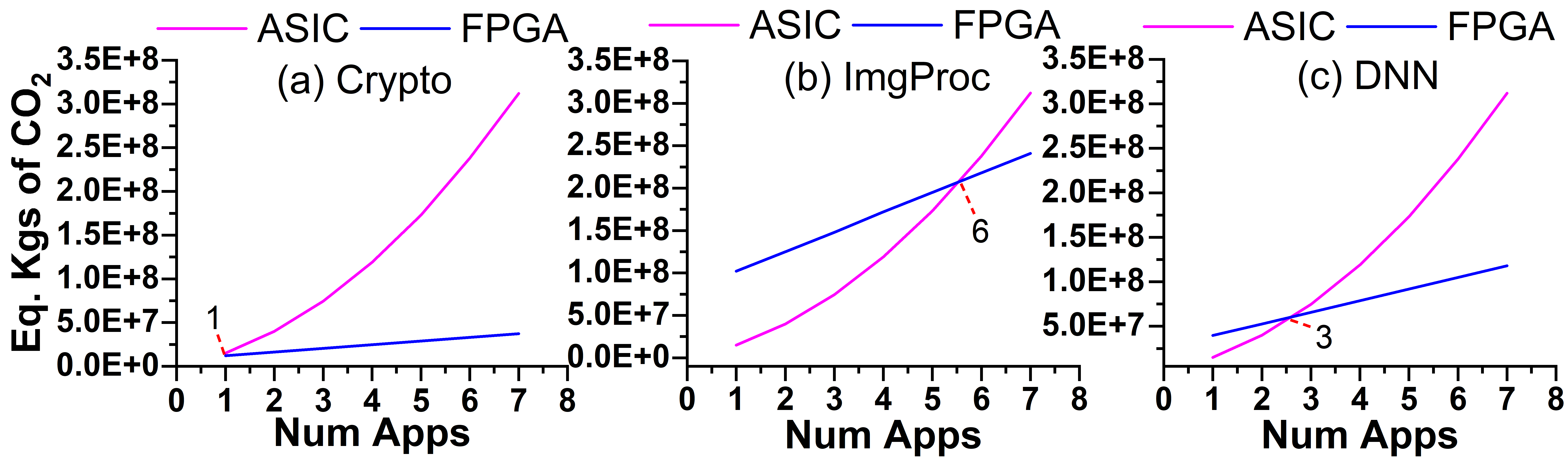}
\vspace{-3mm}
\caption{Comparing CFP of FPGAs and ASICs with $N_\text{app}$; $N_\text{vol}$, $T_\text{i}$, and $f_\text{use}$ are constant.}
\label{fig:cfp-vs-napp}
\vspace{-4mm}
\end{figure}

\begin{figure}[t]
\centering
\includegraphics[width=0.6\linewidth]{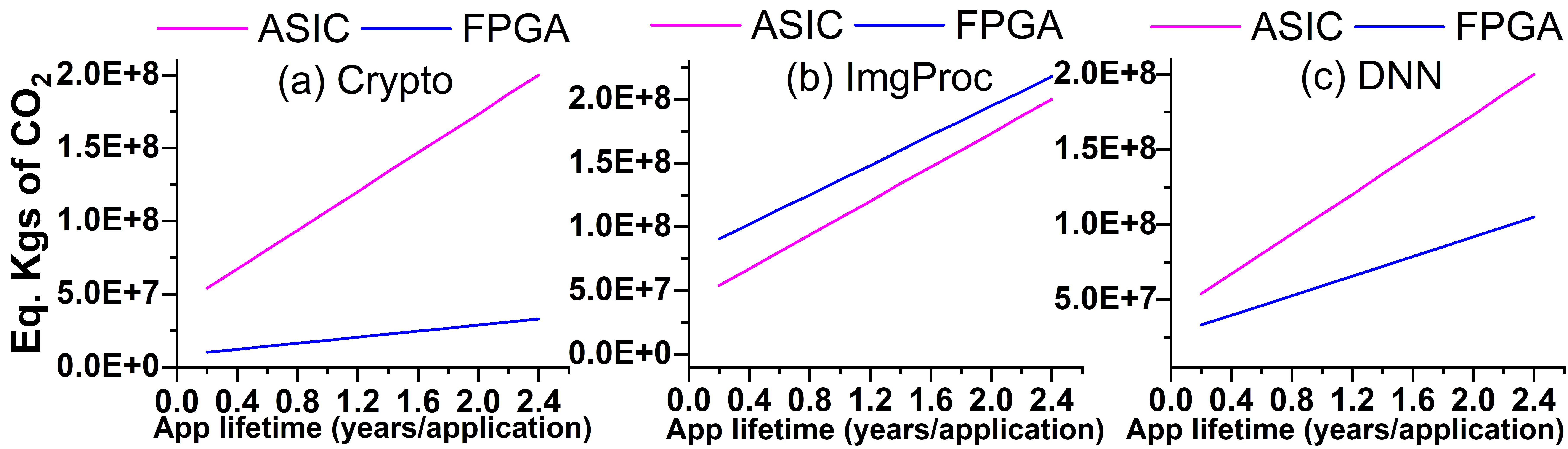}
\vspace{-3mm}
\caption{Comparing CFP of FPGAs and ASICs with $T_\text{i}$; $N_\text{vol}$, $N_\text{app}$, and $f_\text{use}$ are constant.}
\label{fig:cfp-vs-app-lifetime}
\vspace{-6mm}
\end{figure}

\begin{figure}[t]
\centering
\includegraphics[width=0.6\linewidth]{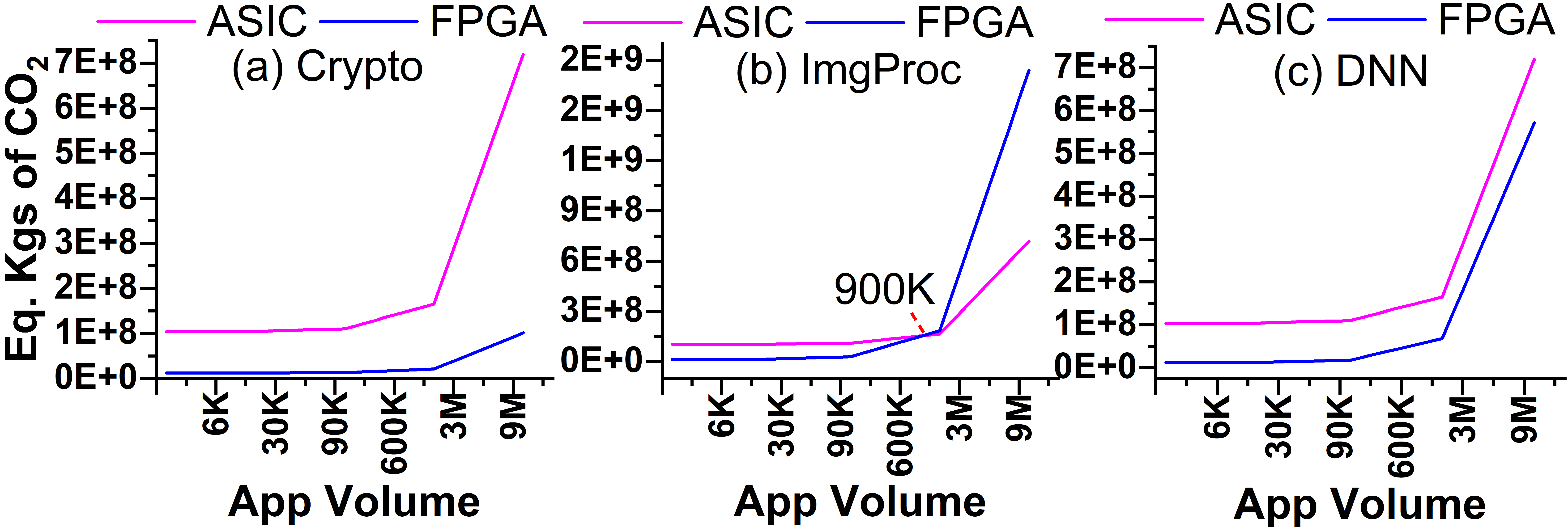}
\caption{Comparing  CFP of FPGAs and ASICs with $N_\text{vol}$; $N_\text{app}, T_\text{i}$, and $f_\text{use}$ are constant.}
\label{fig:cfp-vs-nvol}
\end{figure}

\begin{figure}[t]
\centering
\includegraphics[width=0.6\linewidth]{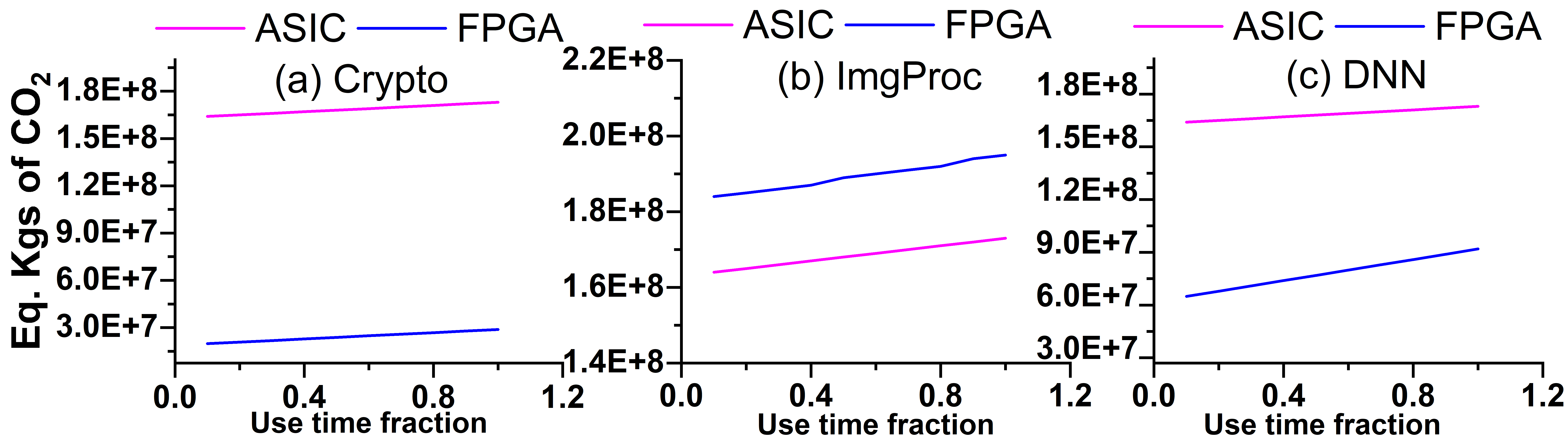}
\vspace{-3mm}
\caption{Comparing  CFP of FPGAs and ASICs with $f_\text{use}$; $N_\text{app}$, $T_\text{i}$, and $N_\text{vol}$ are constant.}
\label{fig:cfp-vs-idletime}
\end{figure}

\noindent\textbf{(A) Impact of number of applications}:
The results of this experiment, where we vary the number of applications from 1 to 8, are shown in Fig.~\ref{fig:cfp-vs-napp}. Notably, different application domains show different behavior, because iso-performance FPGA for each testcase has a different area and power consumption (Table~\ref{tbl:fpga_norm_closed}). 
After the lifetime of an application, a new ASIC must be designed and manufactured to support the next application, whereas the same FPGA can be reconfigured and redeployed across applications. For ASICs in this experiment, the turnover is tied to application lifetime rather than operational lifetime. In addition, the total lifecycle CFP estimated here also includes the impact of EOL-driven device replacement, so the reported trends reflect not only reuse across applications but also the EOL replacement effects.
For Crypto, we observe that the A2F crossover point is achieved after the first application based on the  CFP, because the area and power of the FPGA and ASIC implementations are similar.
For ImgProc, the A2F crossover is at 6 applications, and for DNN, it is at 3 applications for the  CFP. 



\begin{figure}[t]
\centering
\includegraphics[width=0.55\linewidth]{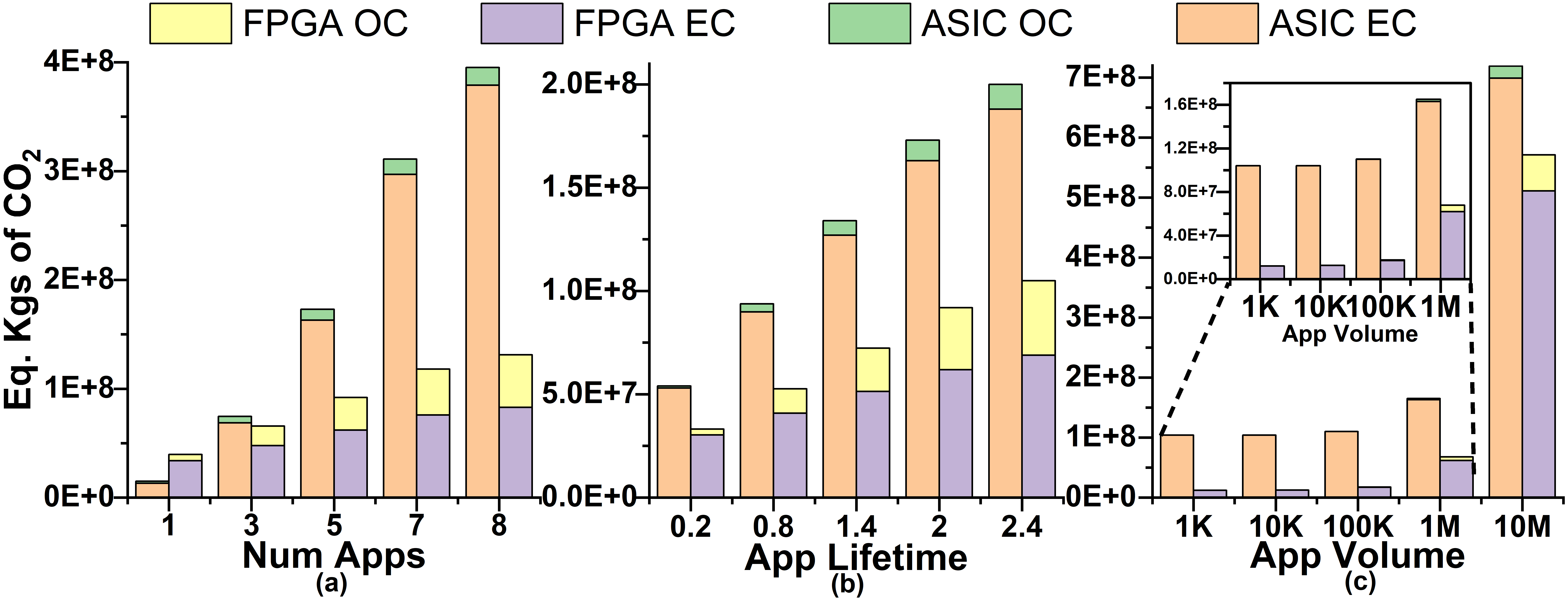}
\vspace{-3mm}
\caption{CFP and its different components for the DNN domain with varying (a) $N_\text{app}$, (b) $T_i$, and (c) $N_\text{vol}$.}
\label{fig:dnn-only}
\end{figure}

\noindent\textbf{(B) Impact of application lifetime}:
The results of this experiment where we vary application lifetime ($T_i$) from 0.2 to 2.4 years are shown in Fig.~\ref{fig:cfp-vs-app-lifetime}.
The CFP increases with application lifetime for all three test cases. For Crypto, FPGA implementations consistently exhibit lower CFP than ASICs across the full range of $T_i$. For ImgProc, ASICs are always more sustainable, irrespective of the application lifetime, due to the large power and area overheads of the FPGA. For DNN testcase, FPGAs are more sustainable across all application lifetimes considered. Unlike the results in~\cite{greenfpga-dac}, where a crossover point was observed, the GreenFPGA framework incorporates a more comprehensive estimation of CFP across the full device lifecycle including expected value and the models for EOL, reliability etc.


\noindent\textbf{(C) Impact of application volume}: 
For this experiment, we vary the volume of each application $N_\text{vol}$ from 1K to 1M.
The results of this experiment are shown in Fig.~\ref{fig:cfp-vs-nvol}.
As the volume increases, the CFP increases as expected.
For Crypto, FPGAs always remain the sustainable option, with the ASIC  CFP being higher even at lower volumes because the iso-performance FPGA for Crypto has similar area and power values compared to the ASIC (Table~\ref{tbl:fpga_norm_closed}), and with the ability to reuse FPGA chips across applications, the CFP for FPGA is lower. For ImgProc the F2A crossover is observed at 900K. For DNNs, FPGAs remain a more sustainable option throughout the full volume range in the experiment. 


\noindent\textbf{(D) Impact of use time fraction}: For this experiment, we vary the use time fraction, $f_\text{use}$ from 0.1 to 0.9. The results of this experiment are shown in Fig.~\ref{fig:cfp-vs-idletime}.
As $f_\text{use}$ increases, the processors are more utilized, increasing the operational carbon and making the operational carbon dominate the total CFP. For Crypto applications, FPGAs are always more sustainable than ASICs for all use time fractions. For image processing tasks, ASICs emerge as the more sustainable option due to their optimized area and power efficiency tailored specifically for such applications across all use time fractions. For the DNN testcase, FPGAs remain more sustainable than ASICs over the full range of use time fractions considered. However, in contrast to Crypto, the FPGA CFP for DNNs has a noticeably steeper slope, suggesting a stronger sensitivity of operational CFP to increasing $f_\text{use}$.

\noindent\textbf{Deep dive into DNN domain's results}: 
In Fig.~\ref{fig:dnn-only}, we show the detailed breakdown of the  CFP of the three experiments listed above (A -- C) for the DNN testcase\footnote{We do not show the breakdown for experiment (D) in the interest of space.}.
We analyze which components dominate the CFP - embodied CFP (EC)  or operational CFP (OC).
When $N_\text{app}$ is varied (Fig. \ref{fig:dnn-only}(a)), the EC of FPGAs increases slightly because the same FPGA is reconfigured and reused across multiple applications, and the increase is primarily due to EOL-driven replacement effects over the extended deployment, but OC increases as the number of applications increases, resulting in greater cumulative operational energy use and hence higher operational CFP. For ASICs, since new ASICs need to be manufactured for each application, EC increases significantly and dominates total CFP.
When $T_i$ is varied (Fig.~\ref{fig:dnn-only}(b)), the EC of both FPGA and ASIC increases. From the OC perspective, comparing to ASICs OC increase, the FPGAs OC contribution grows more rapidly and begins to dominate as the application lifetime increases. This makes ASICs a more favorable choice in terms of operational CFP for longer application lifetimes.

When $N_\text{vol}$ is varied (Fig.~\ref{fig:dnn-only}(c)), the embodied CFP dominates at low volumes, masking the operational CFP contribution. The embodied CFP of ASICs is significantly higher than that of FPGAs, since ASICs cannot be reconfigured across multiple applications. As $N_\text{vol}$ increases, the FPGA embodied CFP also rises, and at larger volumes the difference in total CFP between ASICs and FPGAs becomes much smaller than it is at low volumes.

\begin{figure*}[t]
\centering
\includegraphics[width=0.75\linewidth]{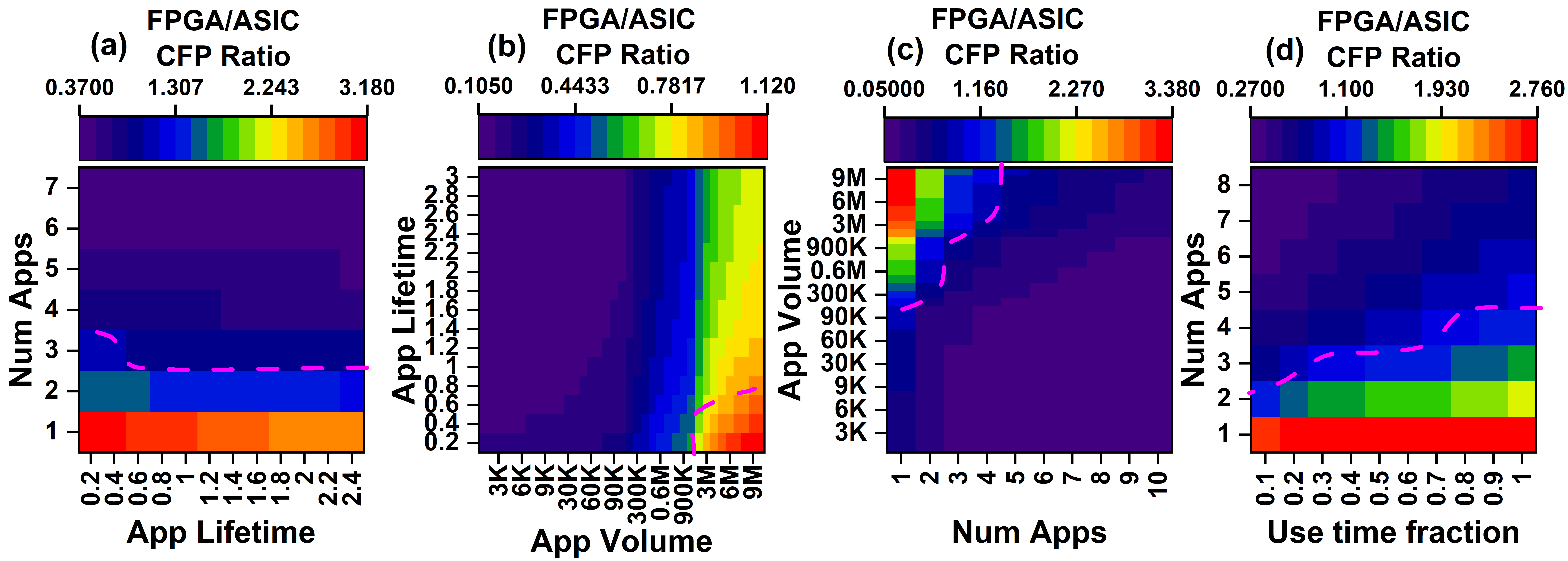}
\vspace{-2mm}
\caption{
Heatmaps showing the ratio of the  CFP of FPGA to ASIC for the DNN application with pairwise sweeps of (a) $N_\text{app}$ and $T_i$ (b), $T_i$ and $N_\text{vol}$, (c) $N_\text{vol}$ and $N_\text{app}$, (d) $N_\text{app}$ and $f_\text{use}$ while keeping the other two variables constant.}
\label{fig:heatmap-sweeps}
\vspace{-7mm}
\end{figure*}

Furthermore, to get more insight into the relationships of the four variables, we perform pairwise sweeps and generate heatmaps.
The results are shown in Fig.~\ref{fig:heatmap-sweeps}.
Each point on the heatmap shows the FPGA to ASIC  CFP ratio.
These heatmaps help understand the regions where ASICs are the more sustainable option (towards red) and where FPGAs are more sustainable (towards blue).
The crossover points are marked using pink dashes (FPGA:ASIC CFP ratio = 1).
For high app volumes ($\approx$9M), FPGAs can be sustainable if number of applications is $>6$.
However, if the volume is high ($>3M$) or the number of applications is low ($<3$), then even lower lifetimes do not make FPGAs sustainable.

Fig.~\ref{fig:heatmap-sweeps} shows the variation of the ratio of the  CFP of FPGA to ASIC for DNN application for pairwise parameter sweeps. 
In Fig.~\ref{fig:heatmap-sweeps}(a), we sweep $T_\text{i}$ and $N_\text{app}$ and plot the ratio of total CFP of FPGAs to ASICs. The pink dashed lines represent the locus where the total CFP of both ASIC and FPGA are equal. Above this locus, FPGAs are a more sustainable option, while ASICs are better below it. Due to the reconfigurability of FPGAs, it is evident that when the number of applications exceeds four, FPGAs are consistently the more sustainable choice compared to ASICs. Fig.~\ref{fig:heatmap-sweeps}(b) presents a heatmap generated by sweeping $T_\text{i}$ and $N_\text{vol}$, with the pink line showing the locus where CFP of ASICs and FPGAs are equal. Since FPGAs have a larger area than ASICs at iso-performance, their overall embodied CFP is higher. This is reflected in the heatmap, FPGAs are the more sustainable option for low-volume applications (below 3M) when the application lifetime is above 0.6 years per application, and they become more sustainable across the full application-volume range for longer lifetimes.
Fig.~\ref{fig:heatmap-sweeps}(c) sweeps $N_\text{app}$ and $N_\text{vol}$ and plots the locus points. In this case, we observe for high volume ($>$90K) with few number of applications ($<$5 $N_\text{app}$) ASIC is more sustainable option, but for all other cases as seen in the image FPGAs are better alternative.
In Fig.~\ref{fig:heatmap-sweeps}(d), we sweep $f_\text{use}$ and $N_\text{app}$ and plot the ratio of CFP of FPGAs to ASICs. The figure and locus indicate that ASICs are generally more sustainable when only a few applications are mapped ($<$3 $N_\text{app}$), especially at higher $f_\text{use}$, whereas FPGAs become the more sustainable option as the number of applications increases.

\subsection {Comparing CFP of FPGAs and GPUs}
\label{sec:cfp-fpga-gpu}
We utilize iso-performance FPGA and GPU from~\cite{eriko-fpga-asic,gpu-fpga-t3} for this experiment and perform analysis for high-performance, energy-efficient, and LHCb target deployments. We define \textit{G2F crossover point}, when the CFP of FPGA becomes lower than GPU, and \textit{F2G crossover point}, when CFP of GPU becomes lower than FPGA. Similar to the previous section,  we evaluate and assess the impact of $N_\text{app}$, $N_\text{vol}$,  $T_\text{i}$, and $f_\text{use}$ on the CFP of FPGAs and GPUs by varying one variable at a time keeping the others constant. For these experiments we use $N_\text{app} = 5$, $N_\text{vol} = 1M$, $T_i = 2$ years, and $f_\text{use} = 0.5$. 
Rapid advances in AI hardware and increasing computation demands lead to increased operational utilization and replacement cycles for GPUs in practice. We use $\lambda_\text{EOL}$ for GPU to be 0.3 $\text{yr}^\text{-1}$ (3.3 years per replacement) and FPGA to be 0.13$\text{yr}^\text{-1}$(7.5 years per replacement)~\cite{greenfpga-dac,altera_fpga_lifetime,obsolete-2,failure-1,refresh-3}.
These parameters are inputs to GreenFPGA and can be adjusted as required for different scenarios.

In addition, unlike ASICs, GPUs can be programmed to support multiple applications on a single GPU. However, new applications may differ significantly to warrant a new GPU to be architected, designed and manufactured.
Therefore, we also sweep the number of applications supported per GPU from two to five for this sensitivity analysis, with GPU\_2App indicating cases where a GPU can be reprogrammed to support two of the applications we consider and GPU\_5App indicating five applications can be supported on the same GPU.

\begin{figure}[t]
\centering
\includegraphics[width=0.95\linewidth]{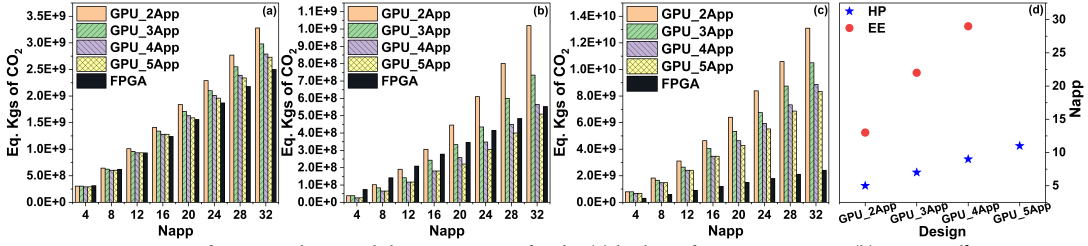}
\vspace{-4mm}
\caption{Comparing CFP of FPGA and GPU while varying $N_\text{app}$ for the (a) high-performance testcase, (b) energy-efficient testcase, (c) LHCb testcase, and (d) G2F crossover point for high-performance and energy-efficient testcase with $N_\text{vol}$, $T_\text{i}$, $f_\text{use}$ being constant.}
\label{fig:fpga-gpu-napp-hp}
\vspace{-6mm}
\end{figure}

\begin{figure}[t]
\centering
\includegraphics[width=0.95\linewidth]{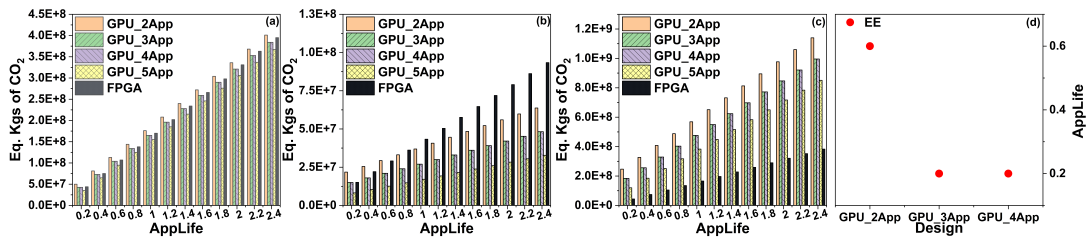}
\caption{Comparing CFP of FPGA and GPU while varying $T_\text{i}$ for (a) high-performance, (b) energy-efficient, (c) LHCb testcase with $N_\text{vol}$, $N_\text{app}$, $f_\text{use}$ constant, (d) F2G crossover point with different numbers of applications per GPU for energy-efficient target deployment.  }
\label{fig:fpga-gpu-applife-hp}

\end{figure}

\noindent\textbf{(A) Impact of number of applications}: We vary $N_\text{app}$ between 1 and 32. 
Depending on the number of applications supported by the GPU, the number of GPUs required for the experiment changes accordingly as we sweep $N_\text{app}$. 
Fig~\ref{fig:fpga-gpu-napp-hp}(a-c) show the variation in CFP with $N_\text{app}$ for different numbers of applications supported per GPU. The corresponding GPU-to-FPGA (G2F) crossover points are summarized in Fig~\ref{fig:fpga-gpu-napp-hp}(d). For the high-performance target deployment shown in Fig~\ref{fig:fpga-gpu-napp-hp}(a), the G2F point shifts from 5 to 11 applications as GPU support increases from GPU\_2App to GPU\_5App, indicating that sharing a GPU across more applications improves its sustainability. For the energy-efficient target deployment in Fig~\ref{fig:fpga-gpu-napp-hp}(b), the G2F point occurs at 13, 22, and 29 applications for GPU\_2App, GPU\_3App, and GPU\_4App, respectively, while no crossover is observed for GPU\_5App within the range considered, the improved power efficiency compared to FPGA helps and delays the G2F point. For the LHCb testcase shown in Fig~\ref{fig:fpga-gpu-napp-hp}(c), no crossover point is observed, as FPGA remains more sustainable than the GPU configurations across the full range of $N_\text{app}$ considered.

\begin{figure}[h]
\centering
\includegraphics[width=0.95\linewidth]{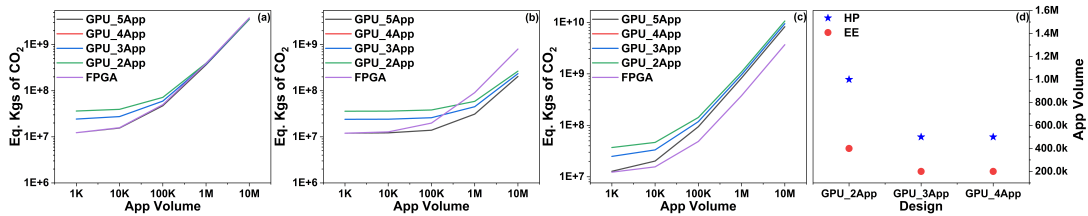}
\caption{Comparing CFP of FPGA and GPU while varying $N_\text{vol}$ for the (a) high-performance, (b) energy-efficient target deployment, (c) LHCb testcase; $N_\text{app}$, $T_\text{i}$, $f_\text{use}$ are constant and (d) shows the F2G crossover point with different number of applications per GPU for high-performance and energy-efficient target deployments.  }
\label{fig:fpga-gpu-vol-hp}
\end{figure}



\noindent\textbf{(B) Impact of application lifetime}: We sweep, $T_i$ from 0.2 to 2.4 years per application.  Fig~\ref{fig:fpga-gpu-applife-hp}(a) shows the variation of CFP with $T_i$ for FPGA and GPUs for the high-performance target deployment. The figure shows that FPGAs are more sustainable than GPUs if the GPUs can be reconfigured for only two applications.  However, for more than two applications, GPU becomes the more sustainable option over the full range of $T_i$ considered. Fig.~\ref{fig:fpga-gpu-applife-hp}(b) shows the corresponding results for the energy-efficient target deployment, while Fig.~\ref{fig:fpga-gpu-applife-hp}(d) summarizes the F2G crossover points. The F2G crossover occurs at 0.6 years per application for the GPU\_2App case, and at 0.2 years per application for the GPU\_3App and GPU\_4App cases. No crossover is observed for GPU\_5App within the plotted range, it shows that with $T_i \geq 0.2$ years per application, GPUs are more sustainable if they can support 3 or 4 applications. Fig.~\ref{fig:fpga-gpu-applife-hp}(c) shows the results for the LHCb testcase, where FPGA remains more sustainable than the GPU configurations for all $T_i$ considered.

\noindent\textbf{(C) Impact of application volume}: We sweep the volume from 1K to 10M and analyze the results for the high-performance, energy-efficient, and LHCb deployments. Fig.~\ref{fig:fpga-gpu-vol-hp}(a-c) show the variation in CFP for FPGA and GPU as $N_\text{vol}$ increases, while Fig.~\ref{fig:fpga-gpu-vol-hp}(d) summarizes the corresponding F2G crossover points for the high-performance and energy-efficient deployments. As the application volume increases, the total CFP also increases due to the greater amount of hardware required, as shown in Fig.~\ref{fig:fpga-gpu-vol-hp}(a-c). The y-axis in these plots is shown in logarithmic scale, making the crossover trends between FPGA and GPU across different reuse scenarios more visible. For the high-performance deployment in Fig.~\ref{fig:fpga-gpu-vol-hp}(a), the FPGA-GPU crossover occurs at approximately \(1\)M applications for GPU\_2App and around \(500\)K applications for GPU\_3App and GPU\_4App. For the energy-efficient deployment in Fig.~\ref{fig:fpga-gpu-vol-hp}(b), the crossover occurs earlier, at roughly \(400\)K applications for GPU\_2App and around \(200\)K applications for GPU\_3App and GPU\_4App, as also summarized in Fig.~\ref{fig:fpga-gpu-vol-hp}(d). This indicates that when GPUs are amortized across more applications, the volume at which they become more sustainable shifts to lower values. For the LHCb testcase in Fig.~\ref{fig:fpga-gpu-vol-hp}(c), no crossover point is observed within the range considered, as FPGA remains the more sustainable option across the full application-volume sweep. We observe that the FPGA-GPU crossover points vary across test cases, reflecting differences in the underlying FPGA and GPU implementations as well as the extent to which the GPU can be amortized across multiple applications.

\begin{figure}[t]
\centering
\includegraphics[width=0.95\linewidth]{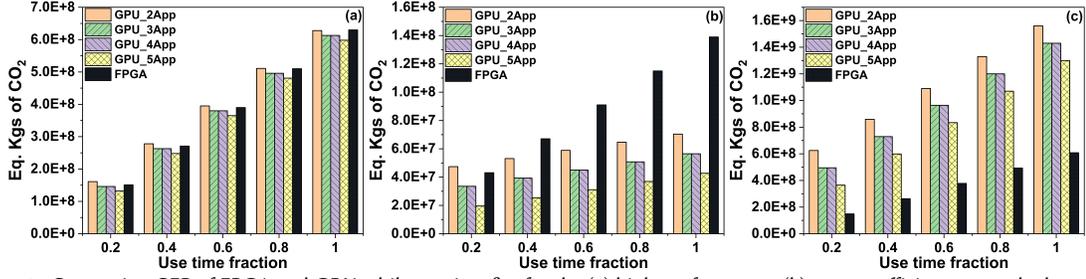}
\vspace{-4mm}
\caption{Comparing  CFP of FPGA and GPU while varying $f_\text{use}$ for the (a) high-performance, (b) energy-efficient target deployment and (c) LHCb testcase; $T_\text{i}$, $N_\text{app}$, $N_\text{vol}$ are constant.}
\label{fig:fpga-gpu-use-time}
\vspace{-4mm}
\end{figure}

\noindent\textbf{(D) Impact of use time fraction}:  We sweep $f_\text{use}$ from 0.1 to 1 while keeping the other three variables, $N_\text{app}$, $N_\text{vol}$, and $T_i$ a constant. Fig.~\ref{fig:fpga-gpu-use-time}(a), (b), and (c) show the results of the experiment for the high-performance, energy-efficient target deployment, and LHCb, respectively. The x-axis shows the use time fraction $f_\text{use}$, and the y-axis shows the total CFP of the FPGA and GPU. Similar to the earlier experiments, we also sweep the number of applications that the GPU can be programmed to support.  For GPU\_2App, from figure the F2G crossover point is at a 90\% use time fraction and 30\% use time fraction for high-performance and energy-efficient deployments, respectively. When the number of applications supported by the GPU is more than two, then there is no crossover point, indicating that for our experiments and test cases, when the GPU can support 3 or more applications, the GPU is more sustainable. We also do not observe any intersection points for LHCb case, with FPGAs being the most sustainable for all $f_\text{use}$.

%

\begin{figure*}[t]
\centering
\begin{minipage}[t]{0.48\textwidth}
    \centering
    \includegraphics[width=\linewidth]{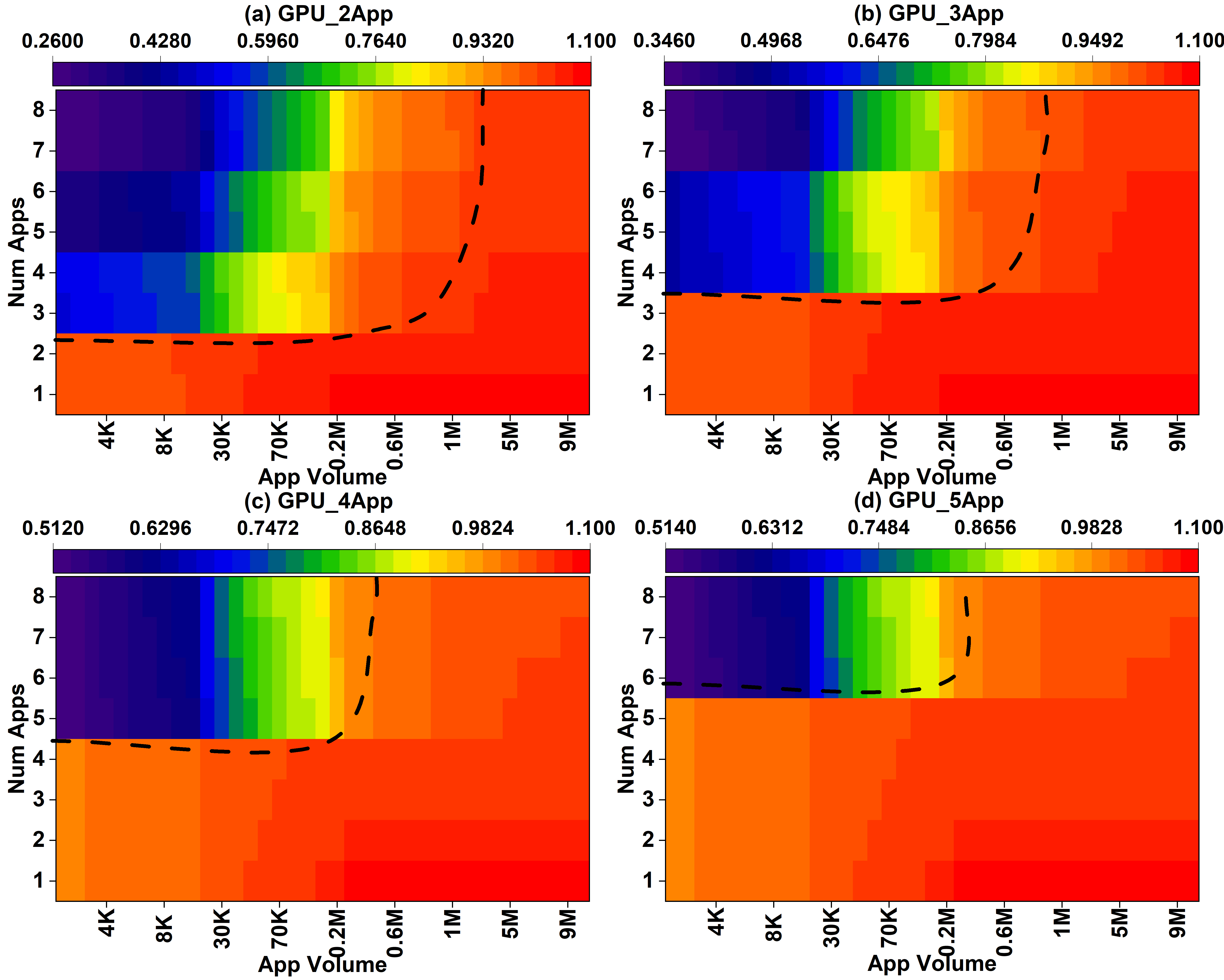}
    \vspace{-5mm}
    \captionof{figure}{Heatmaps of FPGA-to-GPU CFP ratio while sweeping \(N_\text{app}\) and \(N_\text{vol}\) for (a) GPU\_2App, (b) GPU\_3App, (c) GPU\_4App, and (d) GPU\_5App. Other variables are held constant.}
    \label{fig:fpga-gpu-hm-napp-appvol}
\end{minipage}
\hfill
\begin{minipage}[t]{0.48\textwidth}
    \centering
    \includegraphics[width=\linewidth]{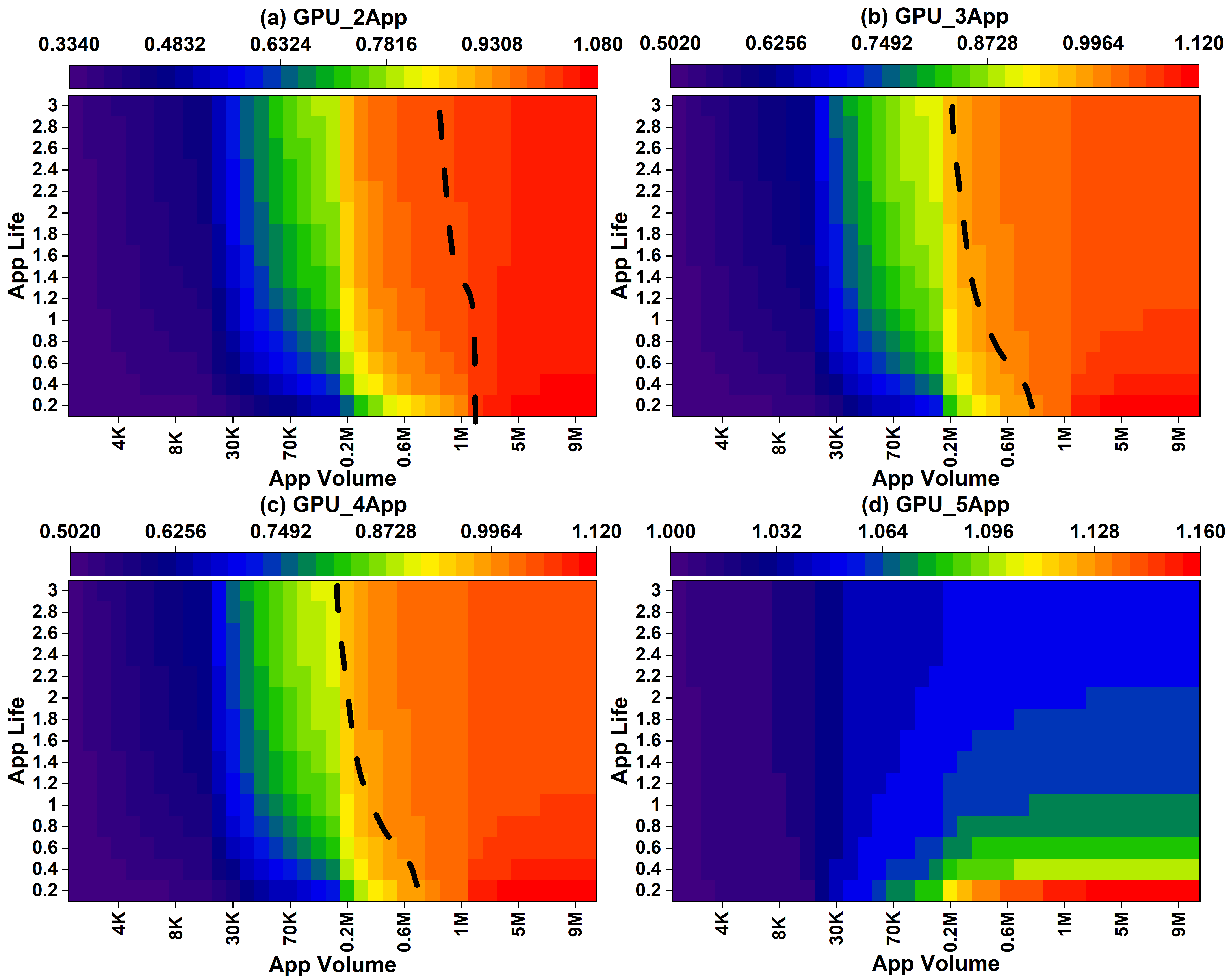}
    \vspace{-5mm}
    \captionof{figure}{Heatmaps of FPGA-to-GPU CFP ratio while sweeping \(T_i\) and \(N_\text{vol}\) for (a) GPU\_2App, (b) GPU\_3App, (c) GPU\_4App, and (d) GPU\_5App. Other variables are held constant.}
    \label{fig:fpga-gpu-hm-applife-appvol}
\end{minipage}
\end{figure*}

%

\begin{figure*}[t]
\centering
\begin{minipage}[t]{0.48\textwidth}
    \centering
    \includegraphics[width=\linewidth]{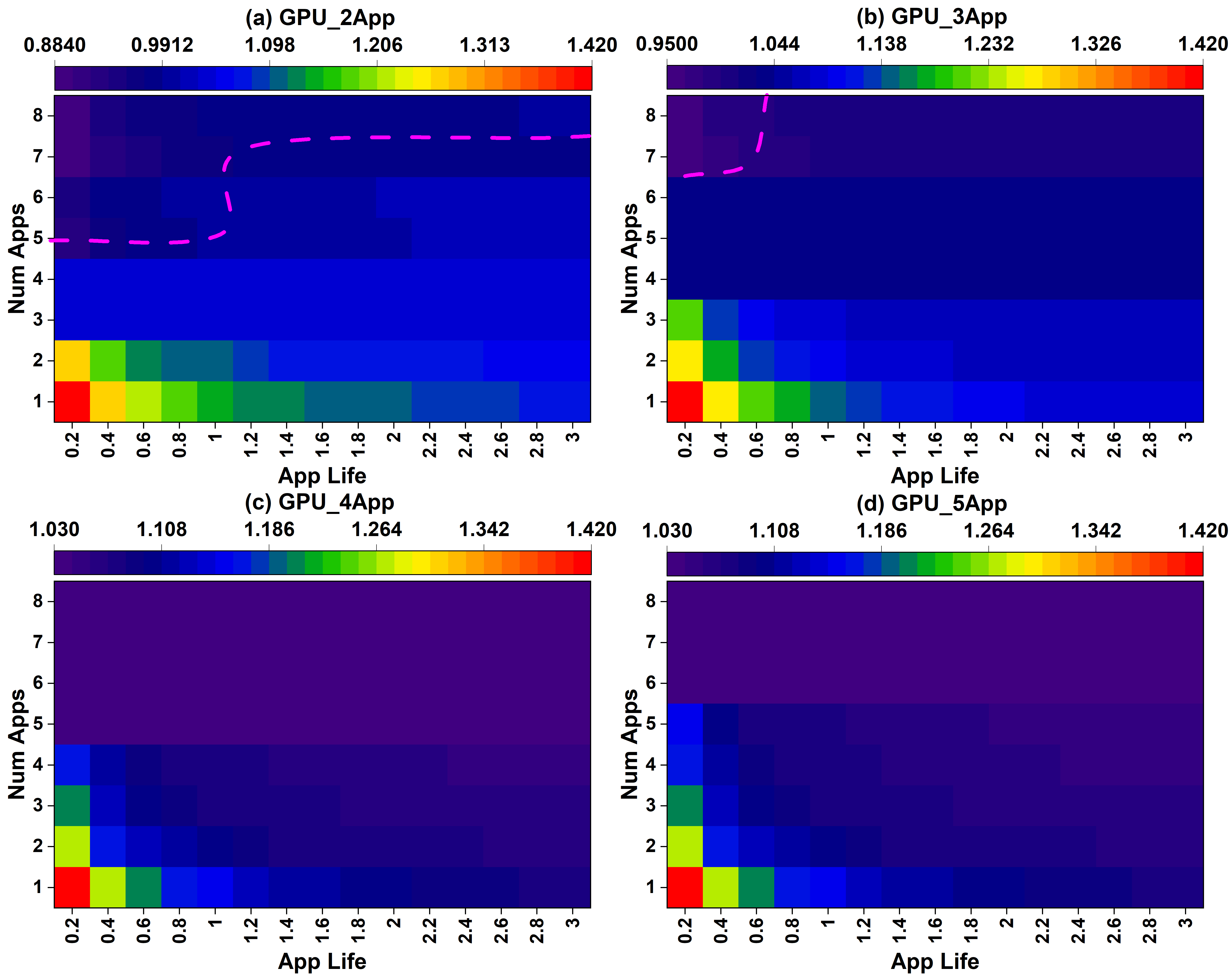}
    \captionof{figure}{Heatmaps of FPGA-to-GPU CFP ratio while sweeping \(T_i\) and \(N_\text{app}\) for (a) GPU\_2App, (b) GPU\_3App, (c) GPU\_4App, and (d) GPU\_5App. Other variables are held constant.}
    \label{fig:fpga-gpu-hm-napp-applife}
\end{minipage}
\hfill
\begin{minipage}[t]{0.48\textwidth}
    \centering
    \includegraphics[width=\linewidth]{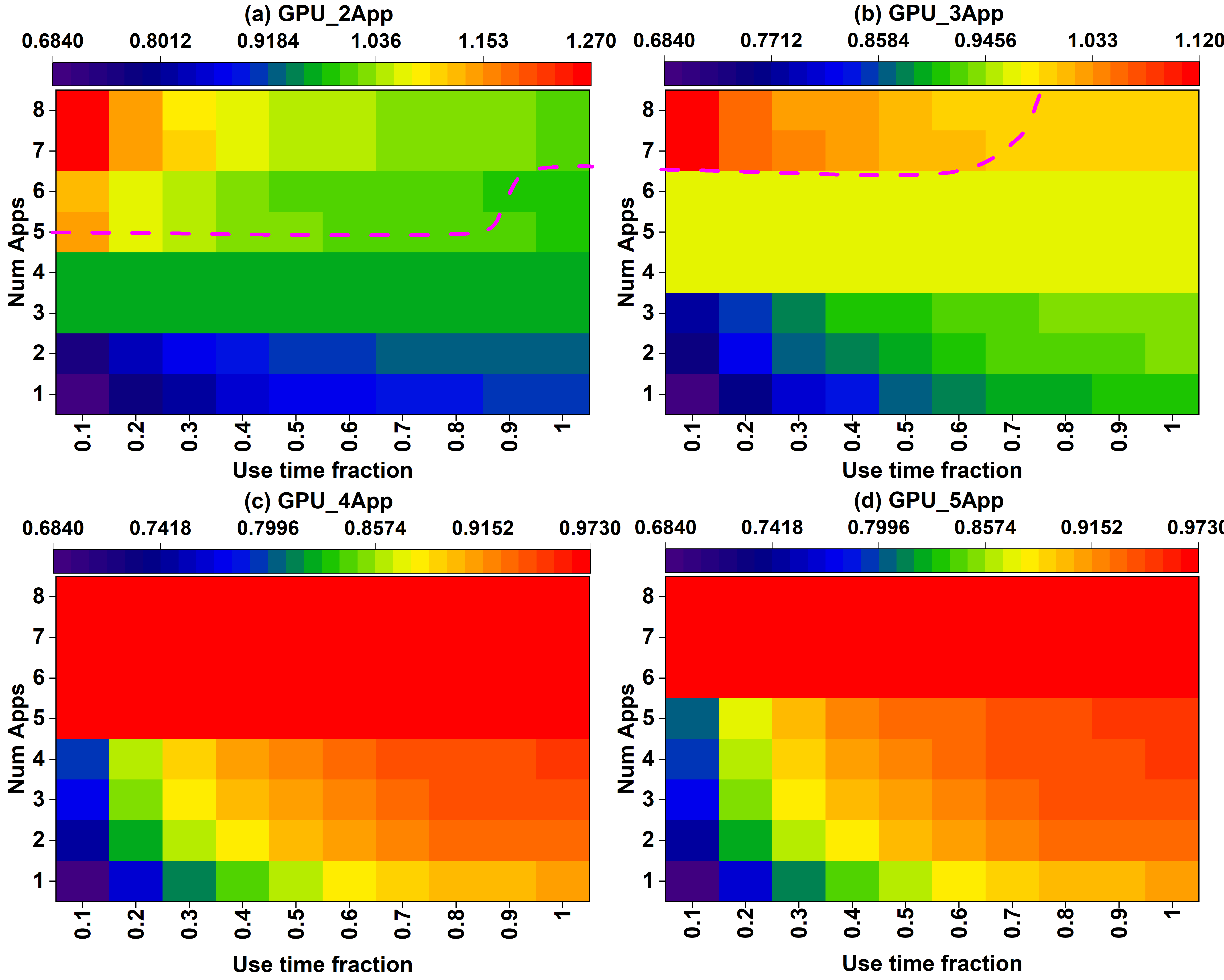}
    \captionof{figure}{Heatmaps of FPGA-to-GPU CFP ratio while sweeping \(f_\text{use}\) and \(N_\text{app}\) for (a) GPU\_2App, (b) GPU\_3App, (c) GPU\_4App, and (d) GPU\_5App. Other variables are held constant.}
    \label{fig:fpga-gpu-hm-usetime}
\end{minipage}
\end{figure*}

\noindent
{\bf Deep dive into high-performance target deployment:} We perform experiments to understand how two-variable sweeps impact our analysis in earlier experiments shown in Section.~\ref{sec:cfp-fpga-gpu}(A -- D). We provide insights into pairwise sweeps of the four variables. The heatmaps in Fig.~\ref{fig:fpga-gpu-hm-napp-appvol} show the ratio of the  CFP of the FPGA to that of the GPU when the GPU can support two to five applications (Fig~\ref{fig:fpga-gpu-hm-napp-appvol}(a -- d)) for $N_\text{app}$ and $N_\text{vol}$ sweeps. The black dashed line indicates the locus where the ratio of the CFPs of FPGA to GPU is equal to 1. Regions above and to the left of the line indicate that FPGAs are more sustainable than GPUs in those scenarios, and regions below the line are where GPUs are more sustainable than FPGAs. Across the plots, we observe a gradual shift in the locus from lower application counts and higher volumes for GPU\_2App toward higher application counts and lower volumes for GPU\_5App.

Fig.~\ref{fig:fpga-gpu-hm-applife-appvol} shows the variation of the ratio of the  CFP of FPGA to that of the GPU when sweeping $N_\text{vol}$ and $T_\text{i}$ for a different number of applications supported by the GPU (a --d). The black dashed line shows the locus of points at which the FPGA's CFP equals that of the GPU. For regions to the left of the figure, FPGAs are more sustainable than GPUs, i.e., for lower application volumes. For the case where the GPU can be configured to support 5 applications, the GPUs are more sustainable for all application volumes and lifetimes, as shown in Fig.~\ref{fig:fpga-gpu-hm-applife-appvol}(d), because the embodied CFP of the GPU can be amortized over its operational lifetime.  

Similarly, Fig.~\ref{fig:fpga-gpu-hm-napp-applife} shows scenarios where $N_\text{app}$ and $T_i$ are swept. FPGA is more sustainable than GPUs in regions above the pink dashed line with low application lifetimes and a higher number of applications (Fig.~\ref{fig:fpga-gpu-hm-napp-applife} (a) and (b)). As the GPU is capable of supporting more applications (Fig.~\ref{fig:fpga-gpu-hm-napp-applife} (c) and (d)), GPUs become more sustainable alternatives to FPGAs. Fig.~\ref{fig:fpga-gpu-hm-usetime} shows scenarios where $f_\text{use}$ and $N_\text{app}$ are swept. FPGA is more sustainable than GPUs in regions above the pink dashed line with a higher number of applications (Fig.~\ref{fig:fpga-gpu-hm-usetime} (a) and (b)).
As the GPU becomes capable of supporting more applications (Fig.~\ref{fig:fpga-gpu-hm-usetime} (c) and (d)), GPUs become more sustainable alternatives compared to FPGAs.

These experiments show that for a large number of applications,  short use times fractions, low volumes, and short application lifetimes, FPGAs can be more sustainable compared to GPUs.

\subsection {Comparing CFP of FPGAs and CPUs}
\label{sec:cfp-fpga-cpu}
\noindent
We will evaluate the impact of applications (Num Apps, $N_\text{app}$), volume of applications (App Volume,  $N_\text{vol}$), use time fraction ($f_\text{use}$), and lifetime of each application (App Lifetime, $T_\text{i}$), of CPUs and FPGAs at iso-performance for a pseudo-random number generation~\cite{cpu-fpga-david}, Llama 2~\cite{cpu-fpga-t2}, and FIR filter~\cite{cpu-fpga-t3} testcases.
In each experiment, we define the \textit{C2F crossover point} as the point when the CFP of the FPGA becomes lower than that of the CPU, and the \textit{F2C crossover point} as the point when the CFP of the CPU drops below that of the FPGA. 
To analyze and compare the performance of CPUs and FPGAs, we use $N_\text{app} = 5$, $f_\text{use} = 0.2$, $T_\text{i} = 2$ years, and $N_\text{vol} = 1M$. We use $\lambda_\text{EOL}$ for CPU to be 0.3 $\text{yr}^\text{-1}$ (3.3 years per replacement) and FPGA to be 0.13$\text{yr}^\text{-1}$(7.5 years per replacement)~\cite{greenfpga-dac,altera_fpga_lifetime,obsolete-2,failure-1,refresh-3}. These parameters are inputs to GreenFPGA and can be adjusted as required for different scenarios.

\begin{figure*}[h]
\centering
\includegraphics[width=0.8\linewidth]{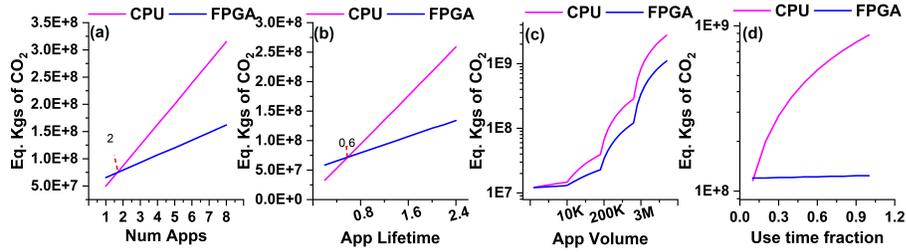}
\caption{Comparing  CFP of FPGAs and CPUs for variations in (a) $N_\text{app}$; $T_\text{i}$, $f_\text{use}$, $N_\text{vol}$ are constant (b) $T_\text{i}$; $N_\text{app}$, $f_\text{use}$, $N_\text{vol}$ are constant (c)
$N_\text{vol}$; $N_\text{app}$, $f_\text{use}$, $T_\text{i}$ are constant (d) $f_\text{use}$; $N_\text{app}$, $T_\text{i}$, $N_\text{vol}$ are constant for random number generation testcase.}
\label{fig:fpga-cpu-napp-15}
\vspace{-4mm}
\end{figure*}

\begin{figure*}[h]
\centering
\includegraphics[width=0.8\linewidth]{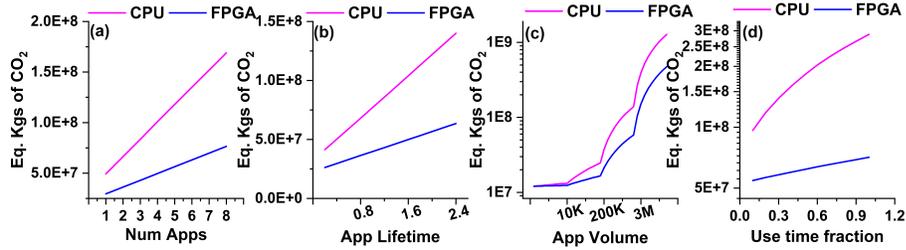}
\caption{Comparing  CFP of FPGAs and CPUs for variations in (a) $N_\text{app}$; $T_\text{i}$, $f_\text{use}$, $N_\text{vol}$ are constant (b) $T_\text{i}$; $N_\text{app}$, $f_\text{use}$, $N_\text{vol}$ are constant (c)
$N_\text{vol}$; $N_\text{app}$, $f_\text{use}$, $T_\text{i}$ are constant (d) $f_\text{use}$; $N_\text{app}$, $T_\text{i}$, $N_\text{vol}$ are constant for Llama 2 testcase.}
\label{fig:fpga-cpu-napp-15-t2}
\end{figure*}

\begin{figure*}[h]
\centering
\includegraphics[width=0.8\linewidth]{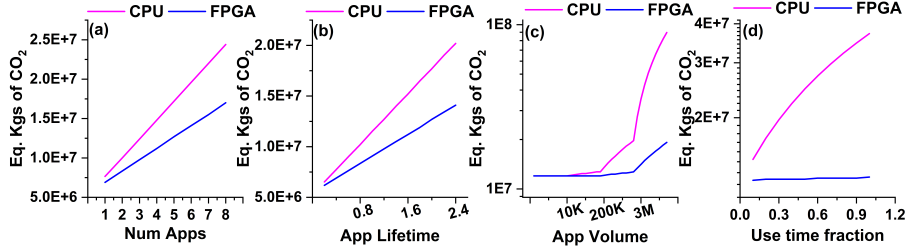}
\vspace{-3mm}
\caption{Comparing  CFP of FPGAs and CPUs for variations in (a) $N_\text{app}$; $T_\text{i}$, $f_\text{use}$, $N_\text{vol}$ are constant (b) $T_\text{i}$; $N_\text{app}$, $f_\text{use}$, $N_\text{vol}$ are constant (c)
$N_\text{vol}$; $N_\text{app}$, $f_\text{use}$, $T_\text{i}$ are constant (d) $f_\text{use}$; $N_\text{app}$, $T_\text{i}$, $N_\text{vol}$ are constant for FIR filter testcase.}
\label{fig:fpga-cpu-napp-15-t3}
\end{figure*}

\begin{figure*}[h]
\centering
\includegraphics[width=0.8\linewidth]{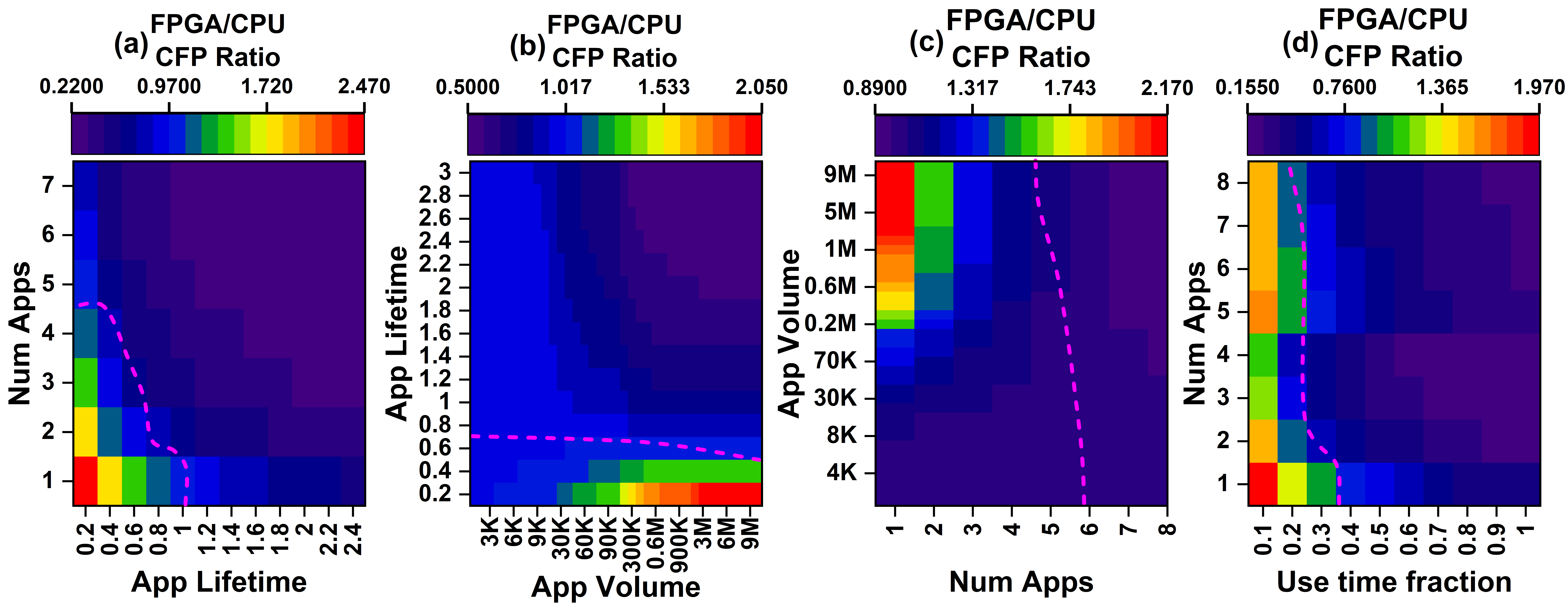}
\vspace{-3mm}
\caption{Comparing the  CFP of FPGA vs. CPU as a ratio with pairwise sweeps of (a) $T_\text{i}$ and $N_\text{app}$, (b) $N_\text{vol}$ and $T_\text{i}$, (c) $N_\text{app}$ and $N_\text{vol}$, and (d) $f_\text{use}$ and $N_\text{app}$ while keeping the other two variables constant.}
\label{fig:fpga-cpu-hm-all}
\vspace{-6mm}
\end{figure*}

\noindent\textbf{(A) Impact of number of applications}: We vary $N_\text{app}$ from 1 to 8. 
For the CPU case study, we assume that the FPGA and CPU require the same number of devices under the iso-performance configuration considered in this analysis (Table~\ref{tbl:fpga_norm_closed}), ensuring a consistent comparison across platforms that can both support a broad range of applications.
Fig.~\ref{fig:fpga-cpu-napp-15}(a) compares the  CFP for CPU with FPGAs as we sweep the number of applications. Since the FPGA has a larger area but is more energy-efficient, we observe a C2F crossover point at 2. The FPGA's embodied CFP is larger than that of the CPU, but as we use it for more applications, this cost is amortized. For other CPU test cases shown in Fig.~\ref{fig:fpga-cpu-napp-15-t2}(a), and Fig.~\ref{fig:fpga-cpu-napp-15-t3}(a), even with more area, since FPGAs are more energy-efficient and have good reconfigurability, for all $N_\text{app}$, they are sustainable alternatives compared to CPU.

\noindent\textbf{(B) Impact of application lifetime}: 
We sweep the application lifetime ($T_\text{i}$) from 0.2 to 2.4 years per application; other variables are held constant. Fig.~\ref{fig:fpga-cpu-napp-15}(b) shows the CFP with the C2F intersection at 0.6 years per application. For the test cases shown in Fig.~\ref{fig:fpga-cpu-napp-15-t2}(b) and Fig.~\ref{fig:fpga-cpu-napp-15-t3}(b), FPGAs are better for all application lifetimes.

\noindent\textbf{(C) Impact of application volume}: Here, volume is swept from 1K to 10M  while keeping other parameters constant. Fig.~\ref{fig:fpga-cpu-napp-15}(c), Fig.~\ref{fig:fpga-cpu-napp-15-t2}(c), and Fig.~\ref{fig:fpga-cpu-napp-15-t3}(c) show CFP with the observation that the FPGAs CFP for all volumes is lower compared to that of the CPUs for all the cases. This is because FPGAs are more power-efficient than CPUs. While CPUs have a lower embodied CFP, the operational CFP for this testcase is the dominant factor, making FPGAs more sustainable.

\noindent\textbf{(D) Impact of usage time}: For this experiment, we sweep $f_\text{use}$ from 0.1 to 1 while keeping the other three variables constant. 
The result for all CPU test cases is presented in Fig.~\ref{fig:fpga-cpu-napp-15}(d), Fig.~\ref{fig:fpga-cpu-napp-15-t2}(d), and Fig.~\ref{fig:fpga-cpu-napp-15-t3}(d).
The variation in CFP with usage time for FPGAs and CPUs demonstrates that FPGAs are more sustainable than CPUs across all usage scenarios. Due to their high reconfigurability, FPGAs can be more efficiently optimized for specific applications compared to CPUs at iso-performance. Since CPUs consume more power at iso-performance, their CFP curve is consistently higher than that of FPGAs. At larger usage times, the operational CFP becomes the dominant contributor to the total CFP, further emphasizing the efficiency advantage of FPGAs.

\noindent
{\bf Deep dive into random number generator testcase}:
We sweep two variables at a time, holding the other two constant, and compare the CFPs of CPUs and FPGAs. Fig.~\ref{fig:fpga-cpu-hm-all}(a) analyzes the variation in CFP with $N_\text{app}$ and $T_\text{i}$. 
The heatmap shows that for $N_\text{app}$ fewer than 5 applications and a per-application lifetime of less than 1 year, CPUs are more sustainable than FPGAs.
All regions to the left of the pink dashed lines are where CPUs are more sustainable than FPGAs, and vice versa on the right. 
Fig.~\ref{fig:fpga-cpu-hm-all}(b) sweeps $N_\text{vol}$ and $T_\text{i}$. The pink dashed line indicates all points above which FPGAs are more sustainable than CPUs. It is clear that for all volumes with application lifetime above 0.7 years per application, FPGAs are better solutions for this testcase than CPUs. 

Fig.~\ref{fig:fpga-cpu-hm-all}(c) sweeps $N_\text{vol}$ and $N_\text{app}$. From the heatmap, it is evident that for all volumes with a lower number of applications ($<$ 6), CPUs are more sustainable, whereas FPGAs are better when we have six or more applications for all application volumes. Regions on the right of the pink dashed lines are where FPGAs are more sustainable than CPUs. Fig.~\ref{fig:fpga-cpu-hm-all}(d) sweeps $N_\text{app}$ and $f_\text{use}$. The locus in pink indicates for lower use time fractions ($\le$ 0.2), CPUs are more sustainable compared to FPGAs for all $N_\text{app}$ range of applications.
The right-hand side of the locus indicates regions where FPGAs are more sustainable than CPUs.

\subsection{Results considering the probabilistic distributions of CFP}
\label{sec:probabilistic-model-results}
\begin{figure}[b]
\centering
\includegraphics[width=0.85\linewidth]{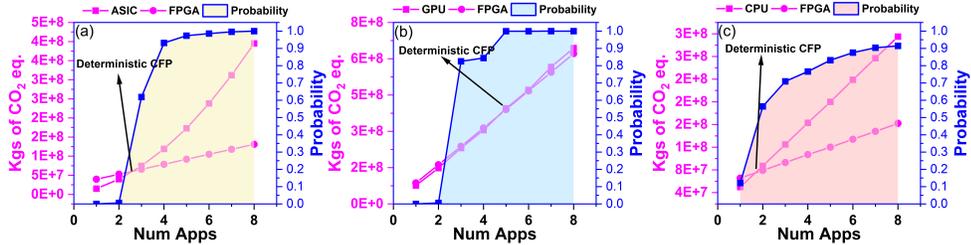}
\vspace{-4mm}
\caption{Comparison of deterministic model~\cite{greenfpga-dac} and probabilistic model (this paper) for (a) ASIC vs. FPGA (DNN testcase) (b) GPU vs. FPGA (high-performance target deployment with GPU\_2App) (c) CPU vs. FPGA (Random Num Gen) for different $N_\text{app}$.}
\label{fig:probability-vertical}

\end{figure}

Results in the previous sections were shown for the expected values of the distributions, here we show how the above experiments are impacted by the probabilistic distribution of CFP. Fig.~\ref{fig:probability-vertical} illustrates the variation in CFP as $N_\text{app}$ is swept while keeping the other three parameters constant for all platforms (ASIC, GPU, and CPU) in comparison with FPGA. The deterministic CFP model from~\cite{greenfpga-dac} evaluates a single CFP value by using the expected values of input parameters. The probabilistic model accounts for inherent uncertainties in these input parameters, yielding a range of CFP values. The figure shows both models - the deterministic CFP is shown in the pink line plot, and the probabilistic CFP is depicted through highlighted regions below the blue line. The left Y-axis represents the CFP values from the deterministic model, while the right Y-axis corresponds to the probability the FPGA is more sustainable than the other computing platform (i.e., the CFP of the FPGA is less than that of the other processor). The pink lines show a single CFP value for each process and a fixed number of applications at which the FPGA becomes more sustainable than the other processor. However, given that CFP is a probabilistic model, there is a probability that the FPGA will become more sustainable than other platforms for a specific number of applications. 

Fig.~\ref{fig:probability-vertical}(a) compares the probabilistic and deterministic models for ASIC vs. FPGA for the DNN application. From the deterministic model, the A2F cross point is at 3 applications, shown at the intersection of the two pink lines, whereas, from the probabilistic model curve shown in the yellow highlighted region, indicates that FPGAs already have a non-zero probability of being more sustainable than ASICs at 2 applications, with this probability increasing sharply from 3 applications onward.
Fig.~\ref{fig:probability-vertical}(b) shows the comparison for GPU vs. FPGA for high-performance target deployment for GPU\_2App. From the deterministic model, the intersection occurs at 5 applications. The blue-highlighted region shows the probability that FPGAs are more sustainable than GPUs, indicating that there is a finite probability that FPGAs are more sustainable than GPUs, even for three applications. 

The comparison for CPU vs. FPGA for the random number generator testcase is shown in Fig.~\ref{fig:probability-vertical}(c), with a deterministic model showing the C2F crossover point to be at 2 applications, and the probabilistic model shows the probability of FPGAs being more sustainable than CPUs for a different number of applications. 
Considering the probabilistic model enables a more comprehensive analysis by accounting for parameter uncertainties. This approach reveals that crossover points may vary and span a range, as depicted in the shaded regions. These variations significantly impact decision-making when evaluating CPUs, FPGAs, ASICs, or GPUs as sustainable alternatives for specific scenarios.

\subsection{Validation and CFP Estimation for Industry Chips}

\begin{figure}[h]
\centering
\includegraphics[width=0.6\linewidth]{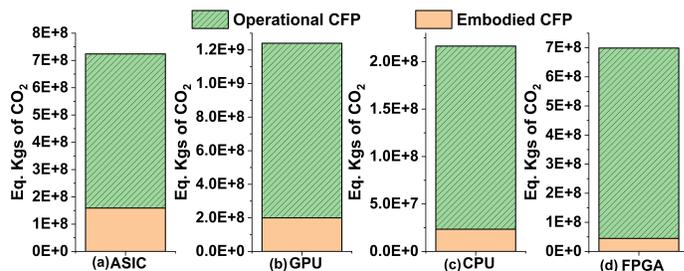}
\vspace{-4mm}
\caption{CFP for industry testcases (a) ASIC, (b) GPU, (c) CPU, and (d) FPGA.} 
\label{fig:industry-cfp}
\end{figure}

All case studies in Fig.~\ref{fig:industry-cfp} use a common setup: a six-year deployment with three applications (each lasting two years) and a total application volume of 1M, using parameters from Table~\ref{tbl:industry-testcases}. Figures (a)–(d) show the estimated CFP for Google TPUv4, NVIDIA H100, Intel Core i9-14900KF, and Intel Agilex 7 FPGA, respectively, with the FPGA reconfigured across applications. Across these cases, operational CFP dominates in all architectures, while embodied contributions vary. The GPU exhibits the highest overall CFP due to its large silicon area and high power consumption. The FPGA shows lower embodied CFP, benefiting from amortization across applications via reconfiguration, although its operational CFP remains significant.

Validating absolute CFP estimates is challenging due to the coarse granularity of publicly available sustainability reports, which often aggregate impacts across entire systems and supply chains rather than isolating chip-level contributions~\cite{mobile-cfp-survey}. Additionally, key parameters for accurate CFP estimation, such as design effort, yield, and detailed manufacturing assumptions, are typically proprietary.

To validate GreenFPGA, we compare its estimates against publicly available reports for TPU and GPU platforms under a consistent setup of three applications over a six-year operational lifetime. For Google TPUv4, GreenFPGA estimates an embodied CFP of 159 kg CO$_2$-eq, compared to 172 kg CO$_2$-eq reported in~\cite{google_sustainability_tpu}. After excluding transportation from the reported value, the difference reduces to 7.6\%. The embodied-versus-operational split is also comparable, with GreenFPGA estimating 22\% embodied and 78\% operational, versus 24.9\% and 75\% in the report. For the NVIDIA H100, GreenFPGA estimates 99 kg CO$_2$-eq per GPU, while the report in~\cite{nvidia_sustainability_h100} implies approximately 109 kg CO$_2$-eq per GPU (derived from an 8-GPU HGX platform), resulting in a 9.8\% difference.

The remaining discrepancies arise from unavailable or proprietary inputs such as yield, transportation, PCB and board-level contributions, manufacturing assumptions, and runtime power profiles. Reported values may also include broader system-level effects that are difficult to isolate at the chip level. Despite these limitations, GreenFPGA captures relative trends and embodied-versus-operational breakdowns with reasonable accuracy. Its open-source framework and configurable parameters enable refinement using higher-fidelity or proprietary data when available, supporting more precise platform-specific CFP estimation.

\section{Conclusion}
\label{sec:conclusion}
\noindent
We propose GreenFPGA, a lifecycle-aware framework for modeling the CFP across the lifecycle of FPGAs, ASICs, CPUs, and GPUs. It integrates probabilistic variations, capturing uncertainties in embodied and operational CFP, including FPGA-specific uncertainty arising from reconfigurability. In addition to refining the model for accurate design-phase CFP estimation, GreenFPGA extends lifecycle modeling with replacement-aware EOL effects due to failure, obsolescence, and refresh/upgrade, reliability-aware operational degradation, and additional embodied CFP components such as testing, and RAM.
Using GreenFPGA, we analyze a range of deployment scenarios to identify sustainable operating regions and compare the CFP trade-offs of different computing platforms.



\bibliographystyle{misc/ieeetr2}
\bibliography{misc/bibfile}

\balance
\flushend

\end{document}